\documentclass[aps,reprint]{revtex4-2}
\usepackage{graphicx} 
\usepackage{xcolor}   
\usepackage{amsmath}  
\usepackage{amsfonts} 
\usepackage{amssymb}  
\usepackage{bm}
\usepackage{bbm}
\usepackage{braket}
\usepackage{tensor}
\usepackage{hyperref}
\usepackage{float}
\usepackage{fullpage}
\usepackage{hhline}
\usepackage{multirow}
\usepackage{array}
\usepackage{hhline}
\usepackage{multirow}
\usepackage{dcolumn}
\usepackage[paperwidth=210mm,paperheight=297mm,centering,hmargin=1.5cm,vmargin=2.5cm]{geometry}
\newcommand\vecbold[1]{\bm{#1}}
\newcommand\im{\mathrm{i}} 

\newcommand\ez{\vecbold{\hat{z}}}

\newcommand\Bpket[1]{\ket{#1}_2}

\newcommand\khat{\vecbold{\hat{k}}}
\newcommand\kvec{\vecbold{k}}
\newcommand\kbar{\vecbold{\bar{k}}}
\newcommand\qvec{\vecbold{q}}
\newcommand\qbar{\vecbold{\bar{q}}}
\newcommand\rvec{\vecbold{r}}
\newcommand\intmeas[1]{\frac{\mathrm{d}^3{\vecbold{#1}}}{|\vecbold{#1}|}}

\newcommand\lambdaBar{\Bar{\lambda}}
\newcommand\sigmabar{\bar{\sigma}}
\newcommand\M{\mathbb{M}}

\newcommand\vecop[1]{\vecbold{{#1}}}

\newcommand\pwket[2]{\ket{\vecbold{#1}\,#2}}

\newcommand\pwbra[2]{\bra{\vecbold{#1}\,#2}}

\newcommand\pw{\pwket{k}{\lambda}}
\newcommand\pwBra{\pwbra{k}{\lambda}}
\newcommand\pwbar{\pwket{\Bar{k}}{\lambdaBar}}
\newcommand\pwBrabar{\pwbra{\Bar{k}}{\lambdaBar}}

\newcommand\biphoton[4]{\pwket{#1}{#2}\pwket{#3}{#4}}

\newcommand\biphotonsimple[2]{\biphoton{#1}{#2}{\Bar{#1}}{\Bar{#2}}}
\newcommand\biphotonsimplebra[2]{\bra{\vecbold{#1}\,#2}\bra{\vecbold{\bar{#1}}\,\bar{#2}}}

\providecommand\Eq[1]{\text{Eq.~(\ref{#1})}}

\begin{document}

\title{A Tensor Product Space for Studying the Interaction of Bipartite States of Light with Nanostructures}

\author{Lukas~Freter$^{\dagger1}$}

\author{Benedikt~Zerulla$^2$}
\author{Marjan~Krsti\'c$^3$}
\author{Christof~Holzer$^3$}
\author{Carsten~Rockstuhl$^{2,3}$}
\author{Ivan~Fernandez-Corbaton$^{*2}$}

\affiliation{{$^1$}Department of Applied Physics, Aalto University School of Science, FI-00076 Aalto, Finland\\
{$^2$}Institute of Nanotechnology, Karlsruhe Institute of Technology (KIT), 76131 Karlsruhe, Germany\\
{$^3$}Institute of Theoretical Solid State Physics, Karlsruhe Institute of Technology (KIT), 76131 Karlsruhe, Germany
}



\begin{abstract}
    Pairs of entangled photons are important for applications in quantum nanophotonics, where their theoretical description must accommodate their bipartite character. Such character is shared at the other end of the intensity range by, for example, the two degenerate instances of the pump field involved in second-harmonic generation. The description and numerical simulation of the interaction of nanophotonic structures with bipartite states of light is challenging regardless of their intensity, and has important technological applications. To address such a challenge, we develop here a theoretical and computational framework for studying the interaction of material structures with bipartite states of light. The theory of the framework rests on the symmetrized tensor product space of two copies of an electromagnetic Hilbert space. For the computational side, the convenient T-matrix method is extended to the tensor product space. When the response of the object to one part of the state is independent of the other part, the T-matrix for bipartite states is a simple function of the typical T-matrix of the single Hilbert space. Such separable material response is relevant, for example, in the interaction of entangled biphoton states with nanostructures. Non-separable operators are identified as the adequate objects to fully integrate non-linear effects such as sum frequency generation or parametric down-conversion. As an example of application, we derive selection rules for second-order non-linear processes in objects with rotational and/or mirror symmetries, and verify them numerically in two different MoS$_2$ clusters.
\end{abstract}

\maketitle
\begin{center}$^\dagger$lukas.freter@aalto.fi,\\$^*$ivan.fernandez-corbaton@kit.edu\end{center}

\section{Introduction}
Within the increasing interest in quantum technologies, one of the objectives of quantum nanophotonics is to exploit the quantum properties of light for applications such as spectroscopy, communication, and sensing \cite{Huang2020,Solntsev2021,Ioannou2021, Oh2022}. Nanophotonic structures can be used to generate quantum states of light, to process and modify them, and ultimately to measure them \cite{Wang2018,Vega2021,Ma2024}. These endeavors raise challenging theoretical questions such as, for example, understanding when and why the entanglement of a biphoton state survives the interaction with a nanostructure \cite{PhysRevLett.121.173901,Lasa-Alonso_2020,tischler2022towards}. However, it is not just at the quantum--level where photons appear in pairs. Considering electromagnetic fields in pairs is also needed in theoretical descriptions of non-linear optical processes, such as sum-frequency generation (SFG) or spontaneous parametric down-conversion (SPDC) \cite{Butcher1990,Boyd2020}. In SFG, the interaction of two incoming fields of different frequencies with matter creates one field with frequency equal to the sum of the frequencies of the incoming fields. The process is called second-harmonic generation (SHG) when the two input fields have the same frequency. SPDC is the process inverse of SFG, where a single field interacts with matter and produces two outgoing fields. The sum of their two frequencies is equal to the frequency of the original single field. The efficiency of these nonlinear processes can be enhanced by nanophotonic structures, such as metasurfaces made from a periodic arrangement of suitably structured individual unit cells \cite{Frizyuk2019,Frizyuk2109b,Sarma2022,Fedotova2023,santiago-cruz_photon_2021,Sharma2023, bonacina_harmonic_2020,verneuil_far-field_2024}.

The intensity of typical entangled photon sources is so low that the quantum regime is reached, where the name biphoton state is fully appropriate. Such a name is, however, not really suitable to refer to the state of the pump for SHG. The two copies of the pump field that are simultaneously involved in the non-linear effect inside the matter have high intensity. We will use the term bipartite state of light, or bipartite field, to refer to the general case including both intensity limits, and reserve the term biphoton state for the quantum regime. 

When dealing with bipartite states in nanophotonics, the theoretical models must describe their interaction with matter without recourse to typical simplifying approximations such as paraxial light or dipolar objects, which are often grossly violated by strongly focused fields, and by the electromagnetic sizes of the objects in question. In some research, symmetries and group theory techniques are being successfully applied to both entanglement evolution and non-linear responses \cite{PhysRevLett.121.173901,Lasa-Alonso_2020,tischler2022towards,Frizyuk2019,Frizyuk2109b}.

For classical fields and single-photon states, symmetries are very conveniently treated in an algebraic approach to light-matter interaction, where the electromagnetic fields belong to a Hilbert space, and a linear operator models the action of a given object on the fields, called the scattering operator of the object, $S$. In particular, symmetry-induced selection rules forbidding specific processes can be derived with just the knowledge of the symmetries of the object, that is, without having the scattering operator at hand. Another advantage of such a setting is that the scalar product of the Hilbert space allows one to readily compute fundamental properties of the field, such as energy and momentum, as well as the transfer of such properties to the object upon light-matter interaction \cite{FerCor2016c}. 

The Hilbert space approach matches very well with a convenient computational tool, the T-matrix, which is a popular and powerful approach to the computation of light-matter interactions for classical fields and single-photon states \cite{Waterman1965,Gouesbet2019,Mishchenko2020}. The T-matrix encodes the full electromagnetic response of a given object upon arbitrary illuminations and can be bijectively mapped to the scattering operator, $S=\mathbbm 1+T$, providing the natural computational implementation of the algebraic approach. Besides describing the response of macroscopic objects characterized by their geometry and material parameters, the T-matrix approach can also be applied to molecules \cite{FerCor2018}. One of the strengths of this approach is that computing the T-matrix of a composite object from the T-matrices of its components is efficient, and extremely efficient in the case of periodic arrangements \cite{Beutel2023b}, such as metamaterials or metasurfaces. T-matrices can be obtained by Maxwell simulations in the case of macroscopic objects, and time-dependent density functional theory (TD-DFT) in the case of molecules. There are many publicly available T-matrix codes \cite{TmatrixCodes}. 

The T-matrix formalism has recently been promoted from its typical monochromatic regime to an inherently polychromatic framework \cite{vavilin_polychromatic_2024}, which allows one to systematically treat light pulses interacting with material objects. The polychromatic T-matrix setting features basis vectors and wavefunctions with well-defined and relatively simple transformation properties under the Poincar\'e group. Additionally, the invariant properties of the electromagnetic scalar product \cite{Gross1964} allow one to consistently define projective measurements in the algebraic setting \cite[Sec.~III]{FerCor2022b}. 

In this article, motivated by the benefits of the algebraic approach, we introduce a framework for the algebraic treatment of the interaction of bipartite states of light with material objects. As an application of the basic formalism, we derive selection rules for non-linear processes, which we verify numerically. We also investigate the extension of the T-matrix to bipartite states. 

The rest of the article is organized as follows. Section~\ref{sec:single} summarizes the most relevant aspects of the electromagnetic Hilbert space $\M$. In Sec.~\ref{sec:biphoton}, we extend such formalism to bipartite states, whose Hilbert space is the tensor product space $\M_2=\M\otimes\M$ restricted by the bosonic permutation rule. It is interesting to note that the extension of theories from a Hilbert space onto the tensor product of two copies of such Hilbert space, which is often known as ``double copy'', has been used in formulations of gravity, analogies between light and gravitation, in particular mapping bipartite states of light to gravitational waves, and in the study of bipartite entanglement, for example \cite{Bern2010,Bern2010b,FerCor2014,Cheung2020,White2021}. We explain in Sec.~\ref{sec:biphoton} how symmetry operators and symmetry generators act in $\M_2$, extend the scalar product from $\M$ to $\M_2$, and derive the condition of entanglement in the bipartite wavefunction. The bipartite wavefunction can encode all the possible biphoton quantum correlations in space, time, and polarization. Section~\ref{sec:scattering} is devoted to light-matter interactions in $\M_2$. We introduce a vacuum state that allows us to promote states in $\M$ to states in $\M_2$. This step is crucial for treating non-linear processes such as SFG or SPDC \cite{Shih_2003,PhysRevA.70.043817}, where single and bipartite states are involved. The formalism built up to this point allows us to easily derive symmetry-induced selection rules for the interaction of bipartite states with material objects. As examples, we obtain selection rules for SFG in objects that feature rotational symmetries and/or mirror symmetries. We show that for objects with discrete rotational symmetry upon a $2\pi/n$ rotation that are illuminated along the symmetry axis with circular polarizations, only the trivial $n=1$ case and $n=3$ case allow SFG in the forward or backward directions, and that, for the $n=3$ case, the two incoming fields need to have the same circular polarization, since otherwise SFG is forbidden by symmetry. We also show that when a mirror symmetric object is illuminated with bipartite fields whose individual linear polarizations are eigenstates of the mirror reflection, the output polarization on any scattering direction contained in the symmetry plane is fixed: transverse magnetic (TM) if the two incoming fields have the same linear polarization (TE--TE or TM--TM), and transverse electric (TE) if they have different polarizations (TE--TM). Generalizations of the selection rules to higher order non-linear processes are briefly discussed. Section~\ref{sec:simulation} contains simulation results that confirm all the tested cases of the selection rules. The simulations of the second-harmonic generation in two different MoS$_2$ clusters are made possible by combining TD-DFT with a solver of Maxwell equations using the Hyper-T-matrix approach \cite{https://doi.org/10.1002/adma.202311405}. In Sec.~\ref{sec:tmatrix}, we investigate the extension of the T-matrix to $\M_2$. We derive the T-matrix for bipartite states when the response of the object is separable, that is when the response to one part of the state {\em does not} depend on the other part. In this case, $T_2$, the T-matrix in $\M_2$, can be obtained from the T-matrix in $\M$:
\begin{equation*}
     T_2=\mathbbm1\otimes T +  T\otimes\mathbbm1 + T\otimes  T,
\end{equation*}
Then, $T_2$ can be used to simulate the change in the entanglement of biphoton states upon interaction with nanostructures, for example. We also discuss how the treatment of non-linear processes requires further research because the separability assumption no longer holds. For example, in SFG, one part of the bipartite state influences the response of the material object to the other part, which indicates the need for non-separable operators. Section~\ref{sec:conc} concludes the article. 

Finally, we would like to highlight the generality of the framework that we introduce in this article. With it, one is able to treat general polychromatic fields, such as short laser pulses, for example. Additionally, the way in which the fields transform under the Poincar\'e group is known, and implemented in publicly accessible code \cite{Vavilin2024b}. These two features can be used, for example, to simulate the pulses of entangled photons used for quantum key distribution in satellite quantum communications \cite{Bourgoin_2013,PhysRevLett_qkd}. Additionally, the equation above allows one to simulate the interaction of such general pulses with nanostructures. Importantly, such simulation only needs the knowledge of the typical T-matrix of the object in $\M$.
\section{Single-photon states\label{sec:single}}
We start by briefly reviewing the theoretical framework for single-photon states and classical fields. See \cite{fernandez-corbaton_helicity_2015,FerCor2022,vavilin_polychromatic_2024} for more details. Let $\M$ denote the Hilbert space of the free field solutions to Maxwell's equations. Then every state $\ket{\phi}\in\M$ satisfying the normalization condition $\braket{\phi|\phi}=1$ can be identified as a \emph{single-photon state} \cite{Gross1964,Zeldovich1965,bialynicki-birula_v_1996}. When $\langle\phi|\phi\rangle=1$, the electromagnetic fields describe a single photon, and when $\langle\phi|\phi\rangle \gg 1$, they describe a classical field. Similarly, the extension to bipartite states will be suitable for describing bipartite beams of very different intensities, such as the very low fluence sources of entangled photons, or the two copies of the laser field involved in SHG.

It is convenient to express arbitrary states in $\M$ as the superposition of a chosen set of basis states, such as, for example, plane waves or spherical waves. The latter are also known as vectors spherical harmonics, or multipolar fields. Within this article, we use the plane-wave expansion 
\begin{equation}
    \ket{\phi} =\sum_{\lambda=\pm1}\int_{\mathbbm{R}^3\setminus \{0\}}\intmeas{k}\phi_\lambda(\kvec)\pw,
    \label{eq:pw_expansion}
\end{equation}
where $\pw$ is a plane-wave state with wave vector $\kvec$ and helicity $\lambda=\pm 1$ (polarization handedness) defined as
\begin{equation}
    \pw = \sqrt{\frac{c\hbar}{\epsilon_0}}\frac{1}{\sqrt2}\frac{1}{\sqrt{(2\pi)^3}}|\kvec|\hat{\vecbold{e}}_\lambda(\khat) \exp(-\im\omega t)\exp(\im\kvec\cdot\rvec),\label{eq:pw}
\end{equation}
where $c$ is the speed of light in vacuum, $\hbar$ the reduced Planck constant, $\epsilon_0$ the vacuum permittivity, $\omega = c|\kvec|$, and $\hat{\vecbold{e}}_\lambda(\khat)$ is the unit polarization handedness vector with $\khat=\kvec/|\kvec|$. The factor $|\kvec|$ in the definition ensures that the plane wave transforms unitarily under the whole Poincar\'e group, including Lorentz boosts \cite{vavilin_polychromatic_2024}. Note that Eq.~\eqref{eq:pw_expansion} yields an integral expression for the space- and time-dependent electric field of the single-photon state $\ket{\phi}$, if one substitutes the plane wave from Eq.~\eqref{eq:pw}. Then, Eq.~\eqref{eq:pw_expansion} is the standard Fourier decomposition of electric fields. From now on, we will drop the summation and integration bounds in the plane-wave expansion and it is understood, that we always sum over both helicities $\pm1$ and integrate over $\kvec\in\mathbb R^3$ excluding the origin. The scalar product in $\M$ can then be written as:
\begin{equation}
\langle \phi|\psi\rangle =\sum_{\lambda}\int\intmeas{k}\phi_\lambda^*(\kvec)\psi_\lambda(\kvec).
\label{eq:scalarproduct}
\end{equation}
The complex expansion coefficient function, or the wave function, shall be given by the scalar product $\braket{\kvec\,\lambda|\phi}=\phi_\lambda(\kvec)$, which fixes the orthogonality relation between plane waves
\begin{equation}
\label{eq:orthogonality}
    \langle \kvec\ \lambda|\kbar\ \lambdaBar\rangle=\delta_{\lambda\lambdaBar}\delta(\kvec-\kbar)|\kvec|.
\end{equation}

In this algebraic setting, transformations such as space-time translations or rotations are unitary linear operators that map states in $\M$ back to $\M$. The scattering of light off an object is described by a linear scattering operator ${S}$, that acts on an incoming state to produce a corresponding outgoing state
\begin{equation}
    \ket{\phi_\text{out}} = S\ket{\phi_\text{in}}.
    \label{eq:S-single}
\end{equation}
Incoming and outgoing states feature incoming and outgoing radiation conditions at infinity, respectively.

\begin{figure}[t!]
\begin{center}
\includegraphics[width = \linewidth]{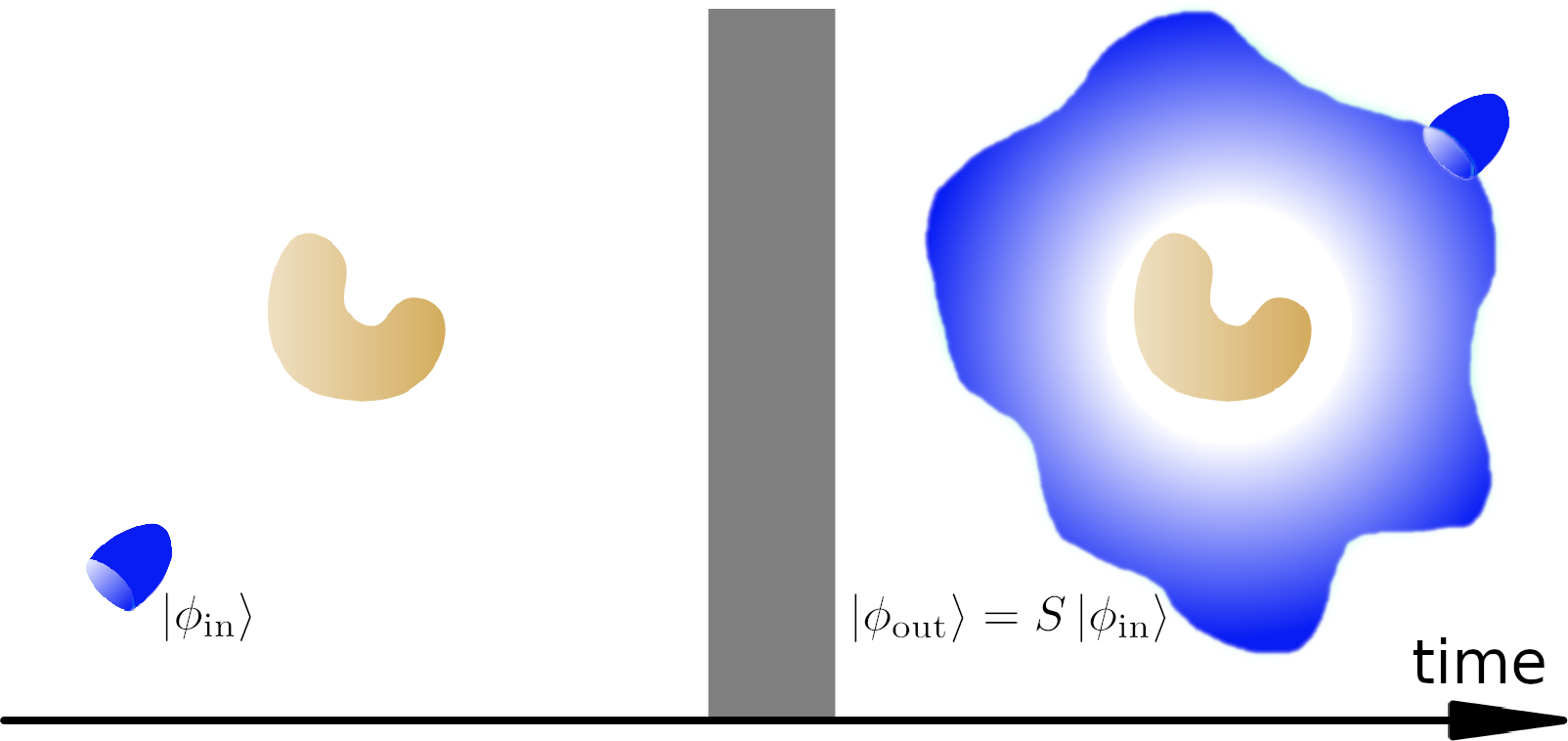}
\end{center}
\caption{Depiction of a typical scattering scenario. Before the light-matter interaction (left), an arbitrary incoming field $\ket{\phi_\mathrm{in}}\in\M$ propagates towards an object described by its scattering operator $S$. $\M$ is the Hilbert space of free solutions of Maxwell equations. After a finite interaction time (gray area), the outgoing field is again a member of $\M$ because we assume that all interactions with the object have subsided. The outgoing field (right) is therefore given by $\ket{\phi_\mathrm{out}}=S\ket{\phi_{\mathrm{in}}}$. The currents induced in the object by the incoming field exist only during the interaction time.}
\label{fig:scattering}
\end{figure}


A sketch of a scattering scenario is depicted in Fig.~\ref{fig:scattering}. On the left, before the scattering process starts, the incoming field $\ket{\phi_\text{in}}\in\M$ propagates towards the object. During the light-matter interaction (gray area in Fig.~\ref{fig:scattering}) the fields are not free, and hence they are not members of $\M$ anymore. However, after all interactions with the object have subsided, the outgoing field $\ket{\phi_\text{out}}$ propagates in free space and is again a member of $\M$. This is the justification of Eq.~\eqref{eq:S-single}: the operator $S$ maps incoming free fields to outgoing free fields. We note that $S$ does not only describe the scattering of light off an object, but also the absorption of light by the object, in which case $S$ is not unitary.

One can represent Eq.~\eqref{eq:S-single} in the plane-wave basis by expanding the input and output states according to Eq.~\eqref{eq:pw_expansion}
\begin{align}
\label{eq:sca_pw_exp1}
    \sum_{\lambda}\int\intmeas{\kvec}\phi_{\text{out},\lambda}(\kvec)\pw = S\sum_{\lambdaBar}\int\intmeas{\kbar}\phi_{\text{in},\lambdaBar}(\kbar)|\kbar\ \lambdaBar\rangle.
\end{align}
After projecting onto the $\pw$ plane wave and making use of the orthogonality of plane waves, we arrive at the intuitive relation
\begin{equation}
\label{eq:sca_pw_exp2}
    \phi_{\text{out},\lambda}(\kvec) = \sum_{\lambdaBar}\int\intmeas{\kbar}\phi_{\text{in},\lambdaBar}(\kbar)S_{\lambda\lambdaBar}(\kvec,\kbar),
\end{equation}
where $S_{\lambda\lambdaBar}(\kvec,\kbar)=\bra{\kvec\,\lambda}S|\kbar\ \lambdaBar\rangle$ is the plane-wave representation of the scattering operator.

Let now the operator $X$ describe a symmetry of the scatterer, for example, a rotation by some angle or a spatial translation by some distance. Then the scattering operator commutes with the symmetry operator:
\begin{equation}
	\label{eq:SXXS}
[S,X]=0\iff S X =  X  S \iff S=XSX^{-1},
\end{equation}
where the inverse exists because $X$ represents a symmetry transformation and must hence be unitary. This simple relation can be used to derive general selection rules. See \cite{fernandez-corbaton_forward_2013} for a specific application.

In some cases, bases other than plane waves are more convenient, such as the basis of multipolar fields. Then, a change of basis allows one to readily obtain the multipolar basis vectors, the multipolar wave function, and the corresponding expression of the scalar product \cite{vavilin_polychromatic_2024}, with which the extension to bipartite states can be carried out following steps parallel to those that we will now take using the plane-wave basis.

We finally note that this formalism applies to general polychromatic solutions of Maxwell's equations, such as short laser pulses, for instance.

\section{Bipartite states and their properties\label{sec:biphoton}}
The formalism introduced in Sec.~\ref{sec:single} is unsuitable to describe light-matter interaction processes where the quantum properties of multiple photons are involved, for example, the interaction of entangled biphoton states with nanostructures. Moreover, the formalism does not cover non-linear processes that change the number of photons and frequency, such as SHG or SPDC.

Describing such processes becomes possible using bipartite states of light, which are naturally built in the tensor product space $\M_2 = \M\otimes\M$ restricted by the bosonic permutation symmetry. In the remainder of this section, we will establish the basic properties of bipartite states.

\subsection{Bipartite state definition and inner product}
A bipartite plane-wave state can be defined as
\begin{align}
    \ket{\kvec\,\lambda,\kbar\,\lambdaBar}_2&=\frac{1}{\sqrt2}(\pw\otimes\pwbar+\pwbar\otimes\pw) , \label{eq:biphoton_pw}
\end{align}
if $\pw\neq\pwbar$, and otherwise as
\begin{equation}
    \ket{\kvec\,\lambda,\kvec\,\lambda}_2 = \pw\otimes\pw.\label{eq:biphoton_pw2}
\end{equation}
Here, subscript 2 indicates that the ket is a member of $\M_2$, and the bosonic permutation symmetry is satisfied due to the plus sign in Eq.~\eqref{eq:biphoton_pw}. A minus sign would satisfy the fermionic case. Therefore, not all possible tensor products of states in $\M$ are valid bipartite states of light. The definition follows the usual convention of symmetrized many-particle states \cite[Sec. 2.1]{Altland_Simons_2006}. For the sake of readability, the tensor product sign "$\otimes$" will often be dropped from now on, as in $\pw\otimes\pwbar \equiv\pw\pwbar$. 
It is worth emphasizing that $\ket{\kvec\,\lambda,\kbar\,\lambdaBar}_2$ is not the same as $\pw\pwbar$ in general, but only if both single-photon plane waves are identical.

One can expand any bipartite state as a superposition of plane waves
\begin{equation}
    \ket{\phi}_2=\sum_{\lambda\lambdaBar}\int\intmeas k\intmeas{\bar k}\phi_{\lambda\lambdaBar}(\kvec,\kbar)\ket{\kvec\,\lambda,\kbar\,\lambdaBar}_2,
    \label{eq:biphoton_pw_expansion}
\end{equation}
where $\phi_{\lambda\lambdaBar}(\kvec,\kbar)$ is any normalizable (wave) function with respect to the scalar product in Eq.~\eqref{eq:scalar_biphoton}. There is a redundancy in the expansion in Eq.~\eqref{eq:biphoton_pw_expansion} in the case of $\pw\neq\pwbar$. Due to the permutation symmetry of the bipartite plane-wave states $\ket{\kvec\,\lambda,\kbar\,\lambdaBar}_2=\ket{\kbar\,\lambdaBar,\kvec\,\lambda}_2$, both $\phi_{\lambda\lambdaBar}(\kvec,\kbar)$ and $\phi_{\lambdaBar\lambda}(\kbar,\kvec)$ contribute to the same state. We may therefore define the symmetrized wave function
\begin{equation}
    \Phi_{\lambda\lambdaBar}(\kvec,\kbar)=\Phi_{\lambdaBar\lambda}(\kbar,\kvec) = \frac{1}{\sqrt2}\left[\phi_{\lambda\lambdaBar}(\kvec,\kbar)+\phi_{\lambdaBar\lambda}(\kbar,\kvec)\right],\label{eq:wf_simple}
\end{equation}
with which Eq.~\eqref{eq:biphoton_pw_expansion} reads
\begin{equation}
    \ket{\phi}_2 = \sum_{\lambda\lambdaBar}\int\intmeas k\intmeas{\bar k}\Phi_{\lambda\lambdaBar}(\kvec,\kbar)\pw\pwbar.
    \label{eq:biphoton_pw_expansion_simple}
\end{equation}
From Eq.~\eqref{eq:biphoton_pw_expansion_simple}, the wave function is given by $\Phi_{\lambda\lambdaBar}(\kvec,\kbar)=\big(\pwBra\pwBrabar\big)\ket{\phi}_2$. If both parts in the bipartite state are identical, the wave function $\phi_{\lambda\lambda}(\kvec,\kvec)$ is already symmetrized, and in analogy to Eq.~\eqref{eq:wf_simple}, it follows that $\Phi_{\lambda\lambda}(\kvec,\kvec) = \phi_{\lambda\lambda}(\kvec,\kvec)$. This relation becomes also clear when comparing the plane-wave expansion in Eq.~\eqref{eq:biphoton_pw_expansion} to Eq.~\eqref{eq:biphoton_pw_expansion_simple} in the case of two identical plane waves in $\M$.

The form of Eq~\eqref{eq:biphoton_pw_expansion_simple} is reminiscent of other expansions of bipartite states, such as \cite[Eq.~(15)]{Shih_2003} for SPDC states. However, Eq.~\eqref{eq:biphoton_pw_expansion_simple} can encode the most general quantum correlations in space, time, and polarization. 

The scalar product between two states in the product space $\M_2$ can be naturally defined using the scalar product in the original space $\M$ as:
\begin{align}
    (\bra{\phi_1}\bra{\phi_2})(\ket{\psi_1}\ket{\psi_2}) = \braket{\phi_1|\psi_1}\braket{\phi_2|\psi_2},
\end{align}
where $\ket{\phi_{1,2}},\ket{\psi_{1,2}}\in \M$.
For two generic bipartite states $\ket{\phi}_2$ and $\ket{\psi}_2$, the scalar product can be straight-forwardly written in the plane-wave basis using Eq.~\eqref{eq:biphoton_pw_expansion_simple} and Eq.~\eqref{eq:orthogonality}:
\begin{align}
\label{eq:scalar_biphoton}
    \tensor[_2]{\braket{\psi|\phi}}{_2}=\sum_{\lambda\lambdaBar}\int\intmeas{k}\intmeas{\bar{k}}\Psi_{\lambda\lambdaBar}^\ast(\kvec,\kbar)\Phi_{\lambda\lambdaBar}(\kvec,\kbar).
\end{align}

We anticipate that the scalar product in $\M_2$ can be used to define projective measurements in $\M_2$ consistently, as done for $\M$ in \cite[Sec.~III]{FerCor2022b}. One of the most important properties of the scalar product in $\M$ is its invariance under the transformations of the conformal group \cite{Gross1964}, which includes the Poincar\'e group. This implies that, for any pair $|\psi\rangle$, $|\phi\rangle$ in $\M$, and $X$ in the conformal group, it holds that $\langle\psi|\phi\rangle=\langle\psi|X^\dagger X|\phi\rangle.$ That is, the scalar product is invariant when the states are transformed with a conformal transformation: $\ket{\psi}\rightarrow X\ket{\psi}$, and $\ket{\phi}\rightarrow X\ket{\phi}$. Since the conformal group in Minkowski space is the largest group of invariance of Maxwell equations \cite{Bateman1910}, the invariance property allows one to define projective measurements by a generic detector mode $|d\rangle$, as $|\langle d|\phi\rangle|^2$, and be assured that their value is the same in all the allowed changes of reference frame \cite[Sec.~III]{FerCor2022b}. The structure of Eq.~\eqref{eq:scalar_biphoton} strongly indicates that a very similar definition should be possible in $\M_2$.

\subsection{Separable and entangled bipartite states\label{sec:sebs}}
A general bipartite state is said to be separable, and hence \emph{not entangled}, if it can be factorized into two independent states in $\M$
\begin{equation}
    \ket{\phi}_2 = \frac{1}{\sqrt 2}\big(\ket{\psi}\ket{\varphi}+\ket{\varphi}\ket{\psi}\big),\quad\ket{\psi},\ket{\varphi}\in\M.
    \label{eq:separability}
\end{equation}
Since $\ket{\psi}$ and $\ket{\varphi}$ are members of $\M$, we can expand them into plane waves according to Eq.~\eqref{eq:pw_expansion}
\begin{align}
    \ket{\psi}&=\sum_\lambda\int\intmeas k\psi_\lambda(\kvec)\pw\label{eq:psi}\\
    \ket{\varphi}&=\sum_{\lambdaBar}\int\intmeas{\bar k}\varphi_{\lambdaBar}(\kbar)\pwbar.\label{eq:phi}
\end{align}
Substituting Eqs.~\eqref{eq:psi} and \eqref{eq:phi} into Eq.~\eqref{eq:separability} yields
\begin{align}
\begin{split}
    \ket{\phi}_2 = &\frac{1}{\sqrt 2}\sum_\lambda\int\intmeas k\psi_\lambda(\kvec)\pw \otimes\sum_{\lambdaBar}\int\intmeas{\bar k}\varphi_{\lambdaBar}(\kbar)\pwbar \\
    + &\frac{1}{\sqrt2}\sum_{\lambdaBar}\int\intmeas{\bar k}\varphi_{\lambdaBar}(\kbar)\pwbar\otimes\sum_\lambda\int\intmeas k\psi_\lambda(\kvec)\pw,
\end{split}
\end{align}
which is easily brought into the following form
\begin{align}
    \ket{\phi}_2 = &\frac{1}{\sqrt 2}\sum_{\lambda\lambdaBar}\int\intmeas k\intmeas{\bar k} \psi_\lambda(\kvec)\varphi_{\lambdaBar}(\kbar)
    \notag\\&\qquad\qquad\times\big(\pw\otimes\pwbar+\pwbar\otimes\pw\big),
\end{align}
and then using Eq.~\eqref{eq:biphoton_pw} we arrive at
\begin{equation}
    \ket{\phi}_2 =\sum_{\lambda\lambdaBar}\int\intmeas k\intmeas{\bar k} \psi_\lambda(\kvec)\varphi_{\lambdaBar}(\kbar)\ket{\kvec\,\lambda,\kbar\,\lambdaBar}_2. \label{eq:separability2}
\end{equation}
Comparing Eq.~\eqref{eq:separability2} with Eq.~\eqref{eq:biphoton_pw_expansion}, we find the foreseeable separability condition for the bipartite wave function 
\begin{equation}
\label{eq:separable}
    \phi_{\lambda\lambdaBar}(\kvec,\kbar) = \psi_\lambda(\kvec)\varphi_{\lambdaBar}(\kbar).
\end{equation}
The $\phi_{\lambda\lambdaBar}(\kvec,\kbar)$ in Eq.~\eqref{eq:biphoton_pw_expansion} does not necessarily meet Eq.~\eqref{eq:separable}, in which case the bipartite state is entangled. Unless specifically mentioned, all the results in the next sections apply to general bipartite states.

\subsection{Symmetry transformations}
Symmetry transformations play an essential role in the algebraic description of electromagnetics because they are intimately linked to conservation laws. The action of common symmetry transformations on bipartite states, such as space-time translations, rotations, duality, parity, and mirror symmetries, can be inferred from their well-known action on states in $\M$. 

Let ${X}$ be an arbitrary symmetry transformation in $\M$. The action of the same transformation in $\M_2$ is then given by \cite[Sec. 3.8]{tung_group_1985}:
\begin{equation}
X_2 = {X}\otimes{X}.
\label{eq:xotimesx}
\end{equation}

As an example, if ${X}$ describes the spatial translation in $\M$ by the displacement vector $\vecbold{s}$, as ${X}=\vecop{T}(\vecbold s)$, we know that
\begin{equation}
    \vecop{T}(\vecbold s)\pw=\exp(-\im\kvec\cdot\vecbold s)\pw.
\end{equation}
Translating a bipartite plane-wave state by $\vecbold s$ is thus achieved by
\begin{align}
    &\vecop T_2(\vecbold s) \ket{\kvec\,\lambda,\kbar\,\lambdaBar}_2\notag\\ &=\frac{1}{\sqrt2}\big(\vecop T(\vecbold s)\pw\otimes\vecop T(\vecbold s)\pwbar + \vecop T(\vecbold s)\pwbar\otimes \vecop T(\vecbold s)\pw \big)\notag\\
    &=\exp[-\im(\kvec+\kbar)\cdot \vecbold s]\ket{\kvec\,\lambda,\kbar\,\lambdaBar}_2,
\end{align}
showing that $\ket{\kvec\,\lambda,\kbar\,\lambdaBar}_2$ is an eigenstate of translations. If the symmetry transformation ${X}$ is continuous, as in the case of spatial translations, it can be expressed by the exponential of a generator $ \Gamma$, namely as $X(\theta)=\exp(-\im\theta\Gamma)$. The generator constitutes the first-order change of the symmetry operator with respect to the continuous parameter $\theta$:
\begin{equation}
    X(\theta)=\mathbbm1-\im\Gamma\theta+\mathcal{O}[\theta^2].\label{eq:generator1}
\end{equation}

Using the above example, the generator of spatial translations is the momentum $\vecop P$, resulting in $\vecop T(\vecbold s) = \exp(-\im \vecop P \cdot \vecbold s/\hbar)$.

To obtain the generator of a symmetry transformation in $\M_2$, we expand the symmetry operator $X_2(\theta)$ to first order in $\theta$. Using Eq.~\eqref{eq:xotimesx} and $X(\theta)=\exp(-\im\theta\Gamma)$, we obtain  
\begin{align}
    X_2(\theta) = \mathbbm1\otimes\mathbbm1 - \im(\Gamma\otimes\mathbbm1+\mathbbm1\otimes\Gamma )\theta + \mathcal{O}[\theta^2],\label{eq:X_2gen}
\end{align}
and the generator can be read off from the linear term (see also \cite[Eq. (7.7-4)]{tung_group_1985})
\begin{equation}
    {\Gamma}_2 =  \Gamma\otimes \mathbbm1 + \mathbbm1 \otimes  \Gamma.
\end{equation}
As is the case in $\M$, if an operator $O_2$ in $\M_2$ commutes with a continuous symmetry $X_2(\theta)=\exp(-\mathrm{i}\theta\Gamma_2)$, then it also commutes with its generator, and the converse statement is also true:
\begin{equation}
O_2X_2(\theta)=X_2(\theta)O_2\iff O_2\Gamma_2=\Gamma_2 O_2.
\end{equation}

It is important to highlight that, in analogy to the case in $\M$ \cite[Chap.~3,\S 9]{Birula1975}, the amount of a fundamental quantity such as energy or momentum contained in a state $\ket{\phi}_2$ can be computed with the scalar product as the ``sandwich'' with the corresponding operator:
\begin{equation}
	\label{eq:stwo}
   \tensor[_2]{\braket{\phi|\Gamma_2|\phi}}{_2}.
\end{equation}

\subsection{Generalization to $\M_N$}
In this section, we briefly discuss how to generalize the formalism to $\M_N=\bigotimes_{n=1}^N\M$, restricted to bosonic permutation symmetry. Formulas for the full states of $N$ bosons are well known, see e.g. \cite[Sec. 2.1]{Altland_Simons_2006}. For example, a triphoton plane-wave state ($N=3$) can be written as
\begin{align}
    \ket{\kvec_1\,\lambda_1,\kvec_2\,\lambda_2,\kvec_3\,\lambda_3}_3
    =&\frac{1}{\sqrt6}(\ket{\kvec_1\,\lambda_1}\ket{\kvec_2\,\lambda_2}\ket{\kvec_3\,\lambda_3}\notag\\
    &+\ket{\kvec_1\,\lambda_1}\ket{\kvec_3\,\lambda_3}\ket{\kvec_2\,\lambda_2}\notag\\
    &+\ket{\kvec_2\,\lambda_2}\ket{\kvec_1\,\lambda_1}\ket{\kvec_3\,\lambda_3}\notag\\
    &+\ket{\kvec_2\,\lambda_2}\ket{\kvec_3\,\lambda_3}\ket{\kvec_1\,\lambda_1}\notag\\
    &+\ket{\kvec_3\,\lambda_3}\ket{\kvec_1\,\lambda_1}\ket{\kvec_2\,\lambda_2}\notag\\
    &+\ket{\kvec_3\,\lambda_3}\ket{\kvec_2\,\lambda_2}\ket{\kvec_1\,\lambda_1}),
\end{align}
if all single-photon states are distinct. All ideas from the previous sections translate to tripartite states. 
\section{Light-matter interaction in $\M_2$\label{sec:scattering}}
\subsection{General setting and emergence of non-linearity}
\label{sec:M2S2}
In a scattering scenario in $\M_2$, we are interested in the fields $\Bpket{\phi_\text{out}}$ after the interaction of some incoming field $\Bpket{\phi_\text{in}}$ with a scattering object, analogous to Fig.~\ref{fig:scattering} in $\M$. Again, we assume that both incoming and outgoing fields are members of $\M_2$, which is valid as long as the incoming field is considered before it enters into contact with the object, and the outgoing field is considered after all the interactions with the scatterer have subsided. In that case, the scattering scenario boils down to a \emph{linear transformation} in $\M_2$, which maps the incoming field to the outgoing field $ S_2:\M_2\rightarrow\M_2$: 
\begin{equation}
    \Bpket{\phi_\text{out}}={S}_2\Bpket{\phi_\text{in}}
    \label{eq:scatter_M2}
\end{equation}
for bipartite states.

Even though the operator ${S}_2$ in \Eq{eq:scatter_M2} is linear in $\M_2$, it can properly model processes that are non-linear in the single field amplitudes. To see how the non-linearity comes about, let us expand incoming and outgoing fields and the scattering operator in the plane-wave basis. After some simple algebra, we arrive at the following equation relating the wave function of the incoming field to the wave function of the outgoing field 
\begin{equation}
    \Phi^\text{out}_{\lambda\lambdaBar}(\kvec,\kbar)=\sum_{\sigma\sigmabar}\int\intmeas{q}\intmeas{\bar{q}} S_{2,\lambda\lambdaBar\sigma\Bar{\sigma}}(\kvec,\kbar,\vecbold{q},\vecbold{\Bar{q}})\Phi^\text{in}_{\sigma\Bar{\sigma}}(\vecbold{q},\vecbold{\bar{q}}),
    \label{eq:biphoton_scatter_coefficients}
\end{equation}
with
\begin{equation}
    \label{eq:biphotonamp}
    S_{2,\lambda\lambdaBar\sigma\Bar{\sigma}}(\kvec,\kbar,\vecbold{q},\vecbold{\Bar{q}}) = (\biphotonsimplebra{k}{\lambda}){S}_2\left(\biphotonsimple{q}{\sigma}\right).
\end{equation}
Here, $\qvec$ and $\bar{\qvec}$ are the wave vectors of the plane waves that expand the incoming bipartite state, and $\sigma$ and $\sigmabar$ are their corresponding helicities. Similarly, $\kvec$, $\kbar$, and $\lambda$ and $\lambdaBar$ are, respectively, the wave vectors and helicities of the plane waves that expand the outgoing bipartite state. 

For illustration purposes, let us now assume that in SHG the incoming field is a separable bipartite state with two equal fields described by the wave function $a_\lambda(\kvec)$. Using Eq.~\eqref{eq:wf_simple} we find that $\Phi^\text{in}_{\sigma\sigmabar}(\qvec,\qbar) = \sqrt2 a_\sigma(\qvec)a_{\sigmabar}(\qbar)$ and $\Phi^\text{in}_{\sigma\sigma}(\qvec,\qvec) = a_\sigma^2(\qvec)$. Equation~\eqref{eq:biphoton_scatter_coefficients} shows that the outgoing wave function depends on the product of two incoming fields, weighted by the appropriate scattering coefficient. Thus, the non-linearity emerges with respect to the single photons or classical fields, whereas the scattering in $\M_2$ is nonetheless described by the linear operator $S_2$.

\subsection{The vacuum state}
In non-linear optical processes such as SFG or SPDC, the number of photons changes, which in the algebraic setting can be seen as going from $\M_2$ to $\M$ or vice versa. However, the scattering operator $ S_2$ maps bipartite states onto bipartite states. For the bipartite formalism to accommodate single states in $\M$, we need to introduce the vacuum state $\ket{0}$. The vacuum state has zero photons and is defined by the following properties
\begin{align}
\begin{split}
    \braket{\kvec\,\lambda|0}&=0,\quad\braket{0|0}=1\\
    \quad{X}(\theta)\ket{0}&=\ket{0},\quad {\Gamma}\ket{0}=0,\quad{X}(\theta)=\exp(-\im\Gamma\theta).\label{eq:vacuum}
\end{split}
\end{align}
Namely, it is a normalized state that is orthogonal to every plane-wave state. It is additionally invariant under any symmetry transformation $ X(\theta)$. This implies that any generator $\Gamma$ generating a symmetry transformation acting on the vacuum yields zero \cite[Sec. 1-4]{Streater1989}. The condition $\braket{\kvec \lambda|0}=0$ implies by linearity that the vacuum state is orthogonal to any state in $\M$, which intuitively makes sense since the probability of finding a state in $\M$ with zero photons is zero. 

The scalar products in the first line of Eq.~(\ref{eq:vacuum}), involving the vacuum state, are different than the scalar product between two members of $\M$ in Eq.~(\ref{eq:scalarproduct}). In particular, $\braket{0|0}=1$ does not imply that the vacuum contains one photon. The vacuum state can be more rigorously built in as the only member of a one-dimensional Hilbert space \cite[Sec.~2.1]{Altland_Simons_2006}, and the following construction in Eq.~\eqref{eq:single_in_bi} can be understood as a tensor product of a plane wave in $\M$ and the lone state in such one-dimensional Hilbert space. For convenience, we nevertheless use the same notation.

We now define the state
\begin{equation}
    \ket{\kvec\,\lambda,0}_2=\pw\ket{0}\label{eq:single_in_bi},
\end{equation}
which we can rigorously identify with the state $\pw$ by checking that when the symmetry transformations of the Poincaré group act on Eq.~\eqref{eq:single_in_bi}, they produce exactly the same effect as on $\pw$. Thus, the vacuum state allows us to write any state in $\M$ as a bipartite state which includes the vacuum $|0\rangle$. 

\subsection{Comparison with second quantization}
The framework described in the previous sections is ultimately equivalent to the formalism of second quantization. There are, however, some conceptual and practical differences. For example, here, the fields are not promoted to operators, and the Hilbert space is not the Fock space. The creation and annihilation operators of second quantization, which provide a very compact formulation of many-body states \cite[Chap.~2]{Altland_Simons_2006}, are absent in our formulation. However, since we are mostly dealing with bipartite states, the notational disadvantage is not very significant. As can be appreciated in the next subsection, the algebraic formalism affords a very direct treatment of symmetries and their consequences, which is arguably more convenient than in second quantization, even though the latter formalism can also be used for the analysis of symmetries \cite{tischler2022towards,Lasa-Alonso_2020,PhysRevLett.121.173901}. The framework that we present here has a more direct connection to computational strategies than second quantization. This can be readily seen in Sec.~\ref{sec:tmatrix}, and also because important physical quantities can be readily computed with the scalar product in $\M_2$. One example is the total amount of a given fundamental quantity such as energy or momentum contained in a given bipartite state, which can be computed through scalar products with \Eq{eq:stwo}. Another example is the consistent definition of measurements, and computation of their outcomes, as discussed right before Sec.~\ref{sec:sebs}.

\subsection{Selection rules for second-order non-linear processes\label{sec:selectionrules}}
In this subsection, we outline several examples of selection rules upon scattering in $\M_2$, reproducing known outcomes, and obtaining some new results. Even though the main focus will be on second-order non-linear processes, we will also comment on generalizations to third- and $N$th-order non-linear processes. 

We assume, that the incoming bipartite state is a separable state of two plane waves in $\M$, and we will explicitly derive the selection rules for SFG. However, the structure of the derivations makes it clear that the selection rules also apply to the inverse SPDC process. Moreover, the procedure for unveiling the selection rules is essentially the same for non-separable states and works for any scattering operator $S_2$ featuring a symmetry.

The SFG process can be exemplified by Eq.~\eqref{eq:scatter_M2} with incoming and outgoing fields given by
\begin{align}
\begin{split}
    \Bpket{\phi_\text{in}} &= \ket{\kvec_1\,\lambda_1,\kvec_2\,\lambda_2}_2,\quad
    \Bpket{\phi_\text{out}}=\ket{\kvec_3\,\lambda_3,0}_2,\\
    \omega_i &= c|\vecbold{k}_i|,\quad i = 1,2,3.
\end{split}
    \label{eq:SFG_in_out}
\end{align}

In the following, we will use several symmetries of the scattering object to infer conservation laws and selection rules. When an object has the symmetry represented by the unitary operator $X$ in $\M$, then the symmetry condition in Eq.~\eqref{eq:SXXS} holds. According to \Eq{eq:xotimesx}, in $\M_2$ the symmetry is represented by the unitary operator $X_2=X\otimes X$. Then the condition
\begin{equation}
	\label{eq:s2sym}
	[S_2,X_2]=0\iff S_2X_2=X_2S_2\iff S_2=X_2 S_2 X_2^{-1}
\end{equation}
follows.

\subsubsection{Time translation symmetry}
Let us first consider the time translation symmetry of the SFG process, which should lead to the preservation of the eigenvalue of the energy operator, that is, the frequency. Time translation symmetry implies that the scattering operator commutes with the generator of time translations, that is, the energy operator $H_2$
\begin{equation}
     S_2 H_2= H_2 S_2.
    \label{eq:symmetry_H}
\end{equation}
Let us apply the energy operator to both incoming and outgoing states in Eq.~\eqref{eq:SFG_in_out}:
\begin{align}
     &H_2\Bpket{\phi_\text{in}}\notag\\& = ( H\otimes\mathbbm 1+\mathbbm 1\otimes H)\frac{1}{\sqrt2}(\ket{\kvec_1\,\lambda_1}\ket{\kvec_2\lambda_2}+\ket{\kvec_2\,\lambda_2}\ket{\kvec_1\lambda_1})\notag\\
    &=\hbar (\omega_1+\omega_2)\ket{\kvec_1\,\lambda_1,\kvec_2\,\lambda_2}_2,
\end{align}
and
\begin{align}
     H_2\Bpket{\phi_\text{out}} &= ( H\otimes\mathbbm 1+\mathbbm 1\otimes H)\ket{\kvec_3\,\lambda_3}\ket{0}\notag\\
    &= \hbar\omega_3\ket{\kvec_3\,\lambda_3,0}_2,
\label{eq:energy_SFG_out1}
\end{align}
where the well known relation for states in $\M$, $ H\pw=\hbar\omega\pw$, is used. With the symmetry relation of Eq.~\eqref{eq:symmetry_H}, we find another expression for the energy of the outgoing state, 
\begin{align}
     H_2 \Bpket{\phi_\text{out}} &=  H_2 S_2\Bpket{\phi_\text{in}} \notag
     \\&= {S}_2 H_2\Bpket{\phi_\text{in}} \notag\\&=  S_2\hbar(\omega_1+\omega_2)\Bpket{\phi_\text{in}}\notag \\&= \hbar(\omega_1+\omega_2)\Bpket{\phi_\text{out}}.
    \label{eq:energy_SFG_out2}
\end{align}
Comparing Eq.~\eqref{eq:energy_SFG_out1} with Eq.~\eqref{eq:energy_SFG_out2} we find the expected relation 
\begin{equation}
	\label{eq:omep}
\omega_3 = \omega_1+\omega_2.
\end{equation}
Equation~\ref{eq:omep} expresses the preservation of the eigenvalue of $H_2$, independently of whether there is absorption by matter, in which case the total energy would not be conserved.

\subsubsection{Rotational symmetric scatterers}
\label{sec:shg_rot}
Let us now consider a discrete rotational symmetry of the scattering object along the propagation direction of the incoming illumination, which can be chosen to be the $z$-direction. The symmetry condition reads
\begin{align}
     S_2 =  R_{z,2}(\theta_n)  S_2  R_{z,2}^{-1}(\theta_n),
    \label{eq:symmetry_R}
\end{align}
with
\begin{equation}
    \theta_n=2\pi/n\text{, } n\in\mathbbm N\text{ and } R_{z,2}(\theta_n) =  R_z(\theta_n)\otimes R_z(\theta_n).
\end{equation}
Such discrete rotational symmetry is often written as $\mathrm{C}_n$ symmetry, as in C\textsubscript{1} (trivial), C\textsubscript{2}, C\textsubscript{3}, and so on.

We will focus on the SFG signal radiated along the $z$-axis, both in forward and backward directions. Then, the incoming and outgoing states read
\begin{equation}
    \Bpket{\phi_\text{in}}= \ket{k_1\ez\,\lambda_1,k_2\ez\,\lambda_2}_2,\quad\Bpket{\phi_\text{out}} = \ket{\pm k_3\ez\,\lambda_3,0}_2,
\end{equation}
where $\ez$ is the unit vector in $z$-direction. The probability for the SFG process to happen is given by the absolute square of the scattering amplitude, which in turn is defined as the matrix element of the scattering operator connecting the input and output states:
\begin{equation}
\label{eq:A}
    A = \tensor[_2]{\braket{\phi_\text{out}| S_2|\phi_\text{in}}}{_2}.
\end{equation}

To study the implications of discrete rotational symmetry on the SFG process, we need to know how the rotation operator $ R_z(\theta_n)$ acts on a state in $\M$ propagating in the $\pm z$ direction. It can be shown that \cite[Sec. 9.7]{tung_group_1985}
\begin{equation}
     R_z(\theta_n) \ket{\pm k\ez\,\lambda} = \exp(\mp\im\lambda\theta_n)\ket{\pm k\ez\,\lambda}.\label{eq:pw_rot}
\end{equation}

To simplify the notation, from now on we will drop the unit vector $\ez$ in the plane-wave states $\ket{\pm k\ez\,\lambda} \equiv \ket{\pm k\,\lambda}$. It is understood that a positive (negative) sign in front of the wave number corresponds to propagation in the positive (negative) $z$-direction.
By making use of the symmetry Eq.~\eqref{eq:symmetry_R}, and relation \eqref{eq:pw_rot}, we can express the scattering \eqref{eq:A} as
\begin{align}
    A&=\tensor[_2]{\braket{\phi_\text{out}| R_{z,2}(\theta_n) S_2 R_{z,2}^{-1}(\theta_n)|\phi_\text{in}}}{_2}\notag\\
    &=\tensor[_2]{\braket{\pm k_3\,\lambda_3,0| R_{z,2}(\theta_n) S_2 R_{z,2}^{-1}(\theta_n)|k_1\,\lambda_1,k_2\,\lambda_2}}{_2}\notag\\
    &=\tensor[_2]{\braket{\pm k_3\,\lambda_3,0| S_2\exp[\im\theta_n(\lambda_1+\lambda_2\mp\lambda_3)]|k_1\,\lambda_1,k_2\,\lambda_2}}{_2}\notag\\
    &= \exp[\im\theta_n(\lambda_1+\lambda_2\mp\lambda_3)]A,
\label{eq:holds}
\end{align}
which tells us that unless the equality $\exp[\im\theta_n(\lambda_1+\lambda_2\mp\lambda_3)] = 1$ holds, only the $A=0$ solution of Eq.~\eqref{eq:holds} exists. For $A=0$, the SFG process is forbidden. This is a selection rule. For a discrete $n$-fold rotational symmetry, we can substitute $\theta_n = 2\pi/n$ yielding the condition
\begin{equation}
    \exp\left[\im\frac{2\pi}{n}(\lambda_1+\lambda_2\mp\lambda_3)\right] = 1.\label{eq:SFG_selectionrule}
\end{equation}
Let us first consider forward SFG, implying a minus sign in Eq.~\eqref{eq:SFG_selectionrule}. There are three distinct cases. First, the two incoming fields can have the same helicity, which is opposite to the helicity of the outgoing state
\begin{equation}
    \lambda_1 = \lambda_2 = -\lambda_3 \Rightarrow \exp\left(\pm\im\frac{6\pi}{n}\right) = 1.
\end{equation}
This is only satisfied for the trivial $n=1$ case and for the $n=3$ case. For all other cases, SFG in the forward direction is forbidden.
Second, the two parts of the incoming state and the outgoing state have the same helicity
\begin{equation}
    \lambda_1=\lambda_2=\lambda_3 \Rightarrow \exp\left(\pm\im\frac{2\pi}{n}\right) = 1,
\end{equation}
resulting in zero forward SFG for all $n>1$. Third, the two incoming fields can have different helicities, resulting in the condition
\begin{equation}
    \lambda_1=-\lambda_2 \Rightarrow \exp\left(\im\lambda_3\frac{2\pi}{n}\right)=1,
\end{equation}
which is again only satisfied for $n=1$.

For backward SFG, the plus sign in Eq.~\eqref{eq:SFG_selectionrule}, we distinguish two cases. The first case is that all fields have the same helicity 
\begin{equation}
    \lambda_1=\lambda_2=\lambda_3\Rightarrow \exp\left(\pm\im\frac{6\pi}{n}\right)=1,
\end{equation}
 which is again satisfied only for $n=1$ and $n=3$. In the second case, 
 one of the fields has a different helicity than the other two, implying that
\begin{equation}
    \exp\left(\pm\im\frac{2\pi}{n}\right)=1,
\end{equation}
which is again only satisfied for $n=1$.

To summarize, except for the trivial non-symmetric case $n=1$, the SFG process in a discrete rotationally symmetric object is only possible for $n=3$ and only if \emph{both incoming fields have the same helicity}. In the forward direction, the outgoing state has the opposite helicity, and in the backward direction, it has the same helicity as the incoming fields. 

\begin{figure*}
    \centering
    \includegraphics[width=\linewidth]{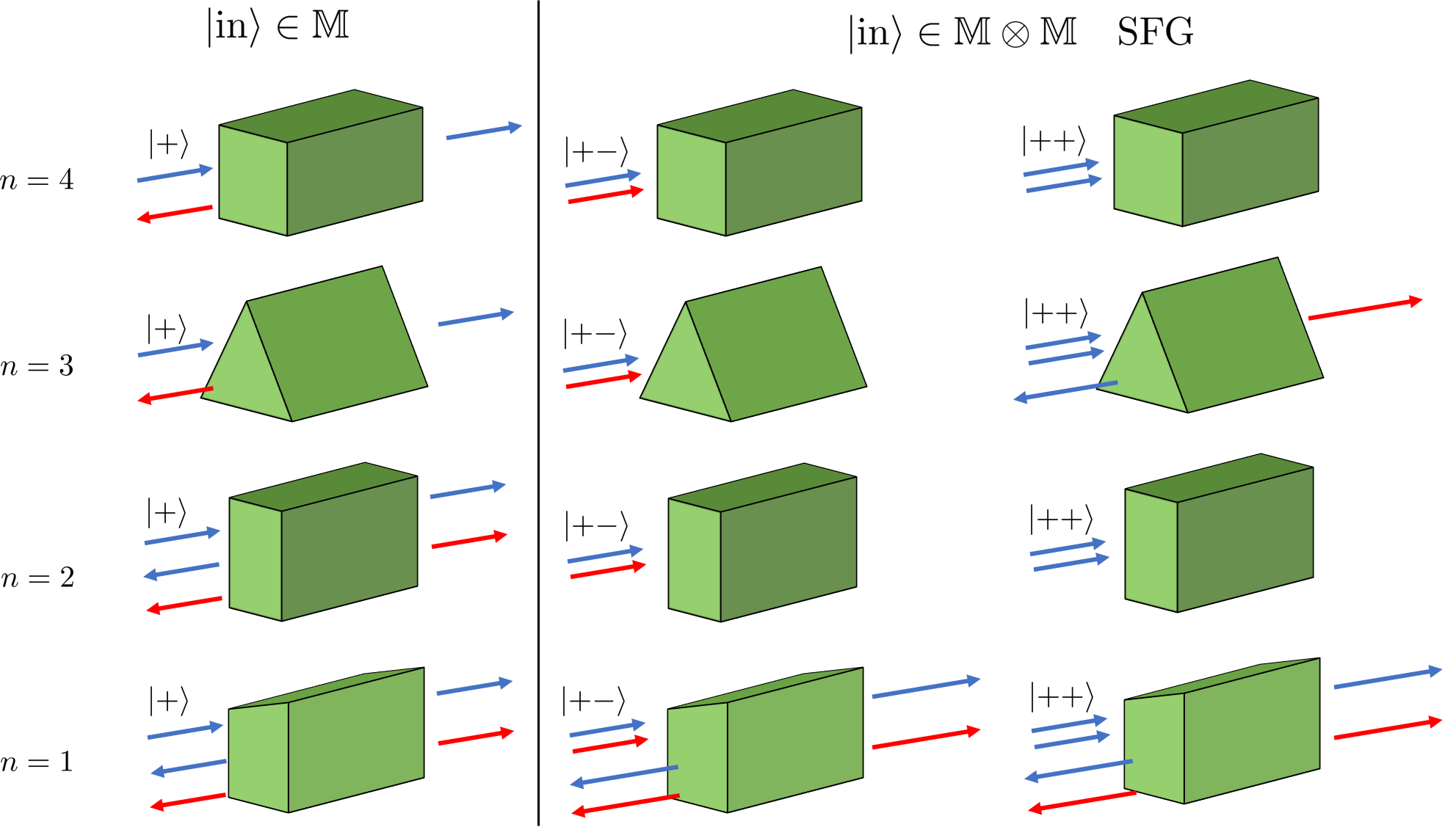}
    \caption{Summary of the selection rules for $n$-fold discrete rotational symmetric scatters. Blue and red arrows indicate left and right polarization handedness, + and - helicity, respectively. Left: Selection rules in $\M$. See Ref.~\cite{fernandez-corbaton_forward_2013} for details. Right: Selection rules for SFG in $\M_2$. The cases with $n>4$ behave as the $n=4$ cases in both $\M$ and $\M_2$.}
    \label{fig:SFG_selection}
\end{figure*}

The selection rules are illustrated in Fig.~\ref{fig:SFG_selection} and compared to the selection rules for linear scattering in $\M$ \cite{fernandez-corbaton_forward_2013}.
Note that the selection rules rely only on the rotational symmetry of the scatterer and not on the apparent mirror symmetries suggested in this figure. So even for a chiral material, for which all mirror symmetries are broken, the selection rules hold.

In \cite{tang_selection_1971, bhagavantam_harmonic_1972}, selection rules for SHG have been developed when both incoming fields have the same helicity, and they agree with the corresponding selection rules derived here. Further, the experimental results in \cite{zhang_chirality_2022} are consistent with the selection rules for $n = 3$. Note, though, that in the supplementary material of the previously mentioned paper, Table S1 is partly incorrect: It indicates the possibility of SFG or SHG for objects with a six-fold rotational symmetry, which we have shown is not possible.

\subsubsection{Generalization to $N$th-harmonic generation}
Generalizing the selection rules due to an $N$th order non-linear process is straightforward under an $n$-fold discrete rotational symmetry. We focus here on the simplest case of $N$th-harmonic generation, where $N$ fields with frequency $\omega = c|\kvec|$ and helicity $\lambda$ interact to produce a single field of frequency $\bar\omega = N\omega = Nc|\kvec|$ and helicity $\lambdaBar$. Assuming propagation along the $z$-direction and again omitting the $\ez$ for the wave vector in the plane-wave kets, the input and output states can be written as
\begin{align}
    \ket{\phi_\text{in}}_N& = \underbrace{\ket{k\,\lambda}\ket{k\,\lambda}\hdots\ket{k\,\lambda}}_{N\text{-times}}\label{eq:in_N}\\
    \ket{\phi_\text{out}}_N &= \ket{\pm N k\,\lambdaBar}\underbrace{\ket0\ket0\hdots\ket0}_{(N-1)\text{-times}}.\label{eq:out_N}
\end{align}
The $N$-part states in Eq.~\eqref{eq:in_N} and Eq.~\eqref{eq:out_N} live in the tensor product space $\M_N = \bigotimes_{i=1}^N\M$ in analogy to the bipartite product space for second-order non-linear processes. Note, that in the case of $N$th harmonic generation, all input photons are equal, meaning we do not need
to take specific care of the bosonic permutation symmetry in \Eq{eq:in_N}.

Following a similar derivation as in the previous section, we arrive at the selection rules for an $n$-fold rotational symmetric medium
\begin{equation}
    \exp\left[\im\frac{2\pi}{n}(N\lambda \mp \lambdaBar)\right]= 1.
\end{equation}
In the case of forward scattering, the condition reads
\begin{align}
    \begin{split}
        &N-1 = mn,\quad m\in\mathbb Z,\text{ for }\lambda = \lambdaBar\\
        &N+1 = mn,\quad m\in\mathbb Z,\text{ for }\lambda = -\lambdaBar,
    \end{split}
\end{align}
which is equivalent to Eq.~(10) in \cite{tang_selection_1971}. It is possible to generalize the above selection rules by allowing e.g. the incoming fields to have different helicities. Then, we would make the substitution $N\lambda \rightarrow \sum_{j=1}^N\lambda_j$, where $\lambda_j$ is the helicity of the $j$-th incoming photon.

\subsubsection{Mirror symmetric scatterers}
\label{sec:mirror_sym}
Let us now derive selection rules for SHG for a medium with mirror symmetry across the $xz$-plane, which is defined by inversion across the $y$-axis ${M}_y:y \rightarrow -y$. Due to the symmetry of the system, it is convenient to switch to the TE/TM basis, defined by linear combinations of circular polarizations as $\ket{\kvec\,\tau} = (\ket{\kvec\,+}+\tau\ket{\kvec\,-})/\sqrt2$, where $\tau=1$ corresponds to TM polarization, and $\tau = -1$ to TE polarization. For the remainder of this section, the wave vector is assumed to be on the $z$-axis and is omitted as an argument of the plane-wave kets. The plane-wave states $\ket{\tau}$ correspond to linearly polarized states where
\begin{equation}
    \ket{\tau = 1} \equiv\ket{\leftrightarrow},\quad\ket{\tau = -1}\equiv\ket{\updownarrow}.
\end{equation}
It is, therefore, easy to see that for states in $\M$, the mirror transformation acts as
\begin{equation}
    {M}_y\ket{\tau} = \tau\ket{\tau}.
\end{equation}
That is, the horizontal (TM, $\hat{x}$) polarization is an eigenstate of the symmetry transformation with eigenvalue $+1$, whereas the vertical (TE, $\hat{y}$) polarization is an eigenstate with eigenvalue $-1$. Horizontal and vertical polarizations are defined with respect to the mirror plane, which is in our case the $xz$-plane. See Fig.~\ref{fig:molecular_models} below representing the exemplary systems that we consider later. 

We readily see that bipartite states with two equal polarization states are always eigenstates of the mirror symmetry with eigenvalue $+1$
\begin{equation}
    {M}_{y,2} \ket{\tau}\ket{\tau} = {M}_y\ket\tau\otimes M_y\ket\tau = \tau^2\ket\tau\ket\tau = \ket\tau\ket\tau.
\end{equation}
To arrive at the selection rule for the SHG process, where two fields with polarization $\tau$ are incoming to produce one field of polarization $\bar\tau$, we compute the scattering amplitude

\begin{align}
    A &= \tensor[_2]{\bra{\bar\tau,0}}{} {S}_2(\ket{\tau}\ket{\tau})\notag\\ &= \tensor[_2]{\bra{\bar\tau,0}}{} {M}_{y,2}{S}_2 {M}_{y,2} (\ket{\tau}\ket{\tau})\notag\\&= \bar\tau A.\label{eq:A_My}
\end{align}
Therefore, the SHG process is only possible for $\bar\tau = 1$, that is, for TM outgoing polarization, independently of the polarization of the incoming fields. Note that if the incoming fields have different linear polarization, the incoming state is an eigenstate of $M_{y,2}$ with eigenvalue $-1$. Proceeding similar to \eqref{eq:A_My}, we find the expression $A=-\bar\tau A$ for the scattering amplitude, where $\bar\tau$ is the polarization of the outgoing photon. The SHG process is then only allowed for TE outgoing polarization. The same derivation and results apply to any scattering direction on the $xz$ plane, for example, backward reflection.

This formalism can be generalized to higher-order processes. For third-harmonic generation, for example, we find that the product of the three incoming polarizations and the outgoing polarization must be equal to one. Hence, if, for example, the three incoming polarizations are equal, then the outgoing polarizations must be the same.

\subsubsection{Tables with the selection rules}
Table~\ref{tab:summary_SHG_rot}, left, shows which output helicities are allowed for a second-order process upon a given incoming helicity when the object has a discrete rotation symmetry about the axis of incidence. Table~\ref{tab:summary_SHG_rot}, right, shows the allowed outgoing linear polarizations of second and third-order processes for an object with $M_y$ mirror symmetry upon linearly polarized illumination.
\begin{table*}
    \centering
    \begin{tabular}{|c|c|c c|}
        \hline
        \multicolumn{4}{|c|}{\textbf{SFG with $\mathbf{C}_n$ symmetry}}\\
        \hline
        $n$&Incident  & Transm./& Refl. \\
        \hline
         1&\multirow{4}{*}{$++$} & $+,-$ & $+,-$\\
         2& & x & x\\
         3& & $-$ & $+$\\
         $\geq4$& & x & x\\
        \hline
        1&\multirow{4}{*}{$+-$} & $+,-$ & $+,-$\\
        2& & x & x\\
        $\geq3$& & x & x\\
        \hline
    \end{tabular}
    \begin{tabular}{|c|c|c|}
        \hline
        \multicolumn{3}{|c|}{\textbf{SFG/ THG with mirror symmetry}}\\
        \hline
        &Incident & Transm./ Refl. \\
        \hline
        \multirow{3}{*}{{SFG}}&TE--TE & TM\\
        &TM--TM & TM \\
        &TE--TM & TE\\
        \hline
        \multirow{4}{*}{{THG}}&TE--TE--TE & TE\\
        &TM--TM--TM & TM \\
        &TE--TE--TM & TM \\
        &TM--TM--TE & TE \\
        \hline
        \end{tabular}
    \caption{Summary of allowed second- and third-order processes in transmission and reflection for an $n$-fold rotational symmetric object (left) and for a mirror symmetric object (right) at normal incidence. An "x" means that the process is forbidden. The allowed and forbidden processes for $--$ and $-+$ incoming helicities can be obtained by flipping all the helicities in the table.}
    \label{tab:summary_SHG_rot}
\end{table*}

\section{Simulation results\label{sec:simulation}}
\begin{figure}[ht]
    \centering
    \includegraphics[width=\linewidth]{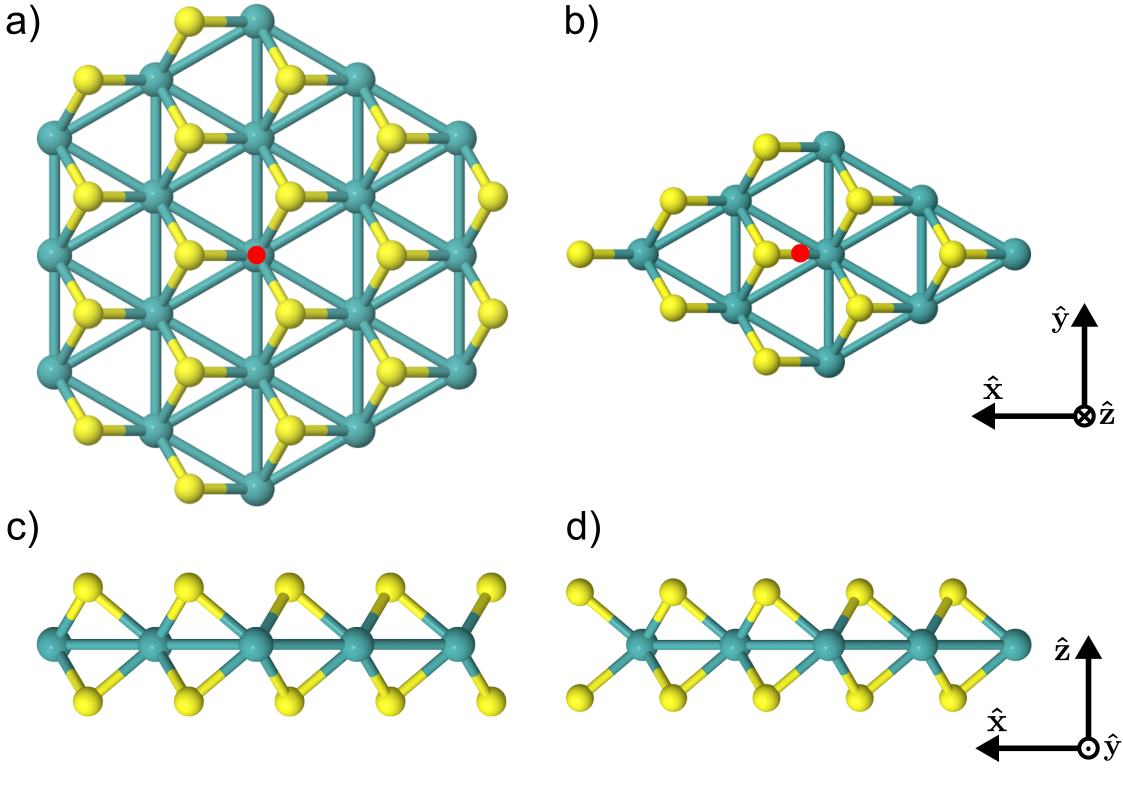}
    \caption{The finite size molecular models of the MoS\textsubscript{2} clusters in \textbf{a)} with C\textsubscript{3} symmetry, and \textbf{b)} with rhomboidal shape and only the trivial C\textsubscript{1} symmetry. Both structures have the mirror symmetry $M_y:\ y\rightarrow -y$. In each case, the center of coordinates was placed at the red dot, which corresponds to the center of mass of the structure. Panels \textbf{c)} and \textbf{d)} show the side view of the hexagon and the rhombus, respectively.}
    \label{fig:molecular_models}
\end{figure}

In this section, we present simulation results confirming the selection rules derived above, specifically for the most interesting case of a three-fold rotation symmetry, that is, $n=3$. We start our simulations from the molecular level by employing quantum chemical calculations based on density functional theory (DFT). Two finite-size molecular models of MoS\textsubscript{2} were chosen. The first one (Fig.~\ref{fig:molecular_models} a)) has an approximately hexagonal shape, but only $R_z(2\pi/3)$ rotational symmetry, and has $M_y$ mirror symmetry. The second molecular model (Fig.~\ref{fig:molecular_models} b)) has a rhomboidal shape, with also $M_y$ mirror symmetry, but does not feature any rotational symmetry along the $z$-axis. We applied time-dependent density functional theory (TD-DFT) calculations to calculate both the linear response, in the form of damped complex dynamic polarizability tensors including electric-electric, electric-magnetic, and magnetic-magnetic components, and the non-linear response, in the form of complex electric-electric hyperpolarizability tensors of these two models at 800 nm and 1064 nm. The Supplementary Information \ref{sec:supplement} contains more details of the simulations. We note that we performed the calculations in a rotated coordinate system due to common point group conventions in the DFT computations. Afterward, we transferred the results to the coordinate system depicted in Fig.~\ref{fig:molecular_models}. Combining the multiple scattering software \emph{treams} \cite{Beutel2023b}  with the Hyper-T-matrix formalism from \cite{https://doi.org/10.1002/adma.202311405}, we can calculate the second-order non-linear response of both systems for different incoming polarizations. Specifically, for both structures shown in Fig.~\ref{fig:molecular_models}, we simulated the SHG response for normally incoming fields at the fundamental wavelength. 

In the following, we discuss the simulation results for the fundamental wavelength of 800 nm. We note that the qualitative result is the same for 1064 nm. For the rhomboid structure in Fig.~\ref{fig:molecular_models} b),  the intensity of the SHG wave in transmission and reflection for TE and TM polarized incoming and outgoing fields is shown in Table~\ref{tab:rhomboid_800nm_lin}. We note that we only evaluate outgoing plane waves with wave vectors on the $z$-axis. We further note that all of our simulation results are normalized to the largest intensity, which occurred for two TM-polarized incoming fields in TM transmission and reflection. Then, normalized signals on the order of 10$^{-5}$ and lower are considered numerical noise. The value 1 corresponds to an SHG conversion efficiency equal to 6.1$\times 10^{-34}$, which is very low because of the very small sizes of the clusters. As seen from the results in Table~\ref{tab:rhomboid_800nm_lin}, if the two incoming fields have the same linear polarization, then the SHG signal in transmission and reflection will only be in the TM polarization. On the contrary, if the incoming fields have different linear polarizations, only TE polarization appears in the SHG signal. This is fully consistent with the selection rules derived in Section~\ref{sec:mirror_sym}. The different outcomes for different linear polarizations of the incoming fields are consistent with the fact that the rhomboid structure is anisotropic, that is, the $x$ and $y$ axes are not equivalent in the rhomboid.

Table~\ref{tab:rhomboid_800nm_circ} shows the simulated SHG intensity for circularly polarized fields incident on the rhomboid structure. The first row shows the response where two fields at the fundamental wavelength of the same helicity, namely plus helicity, combine to give rise to a field at the second-harmonic wavelength. The results for two incoming fields with different helicities are shown in the second row. Due to the trivial C\textsubscript{1}-symmetry, we expect SHG signals both in reflection and transmission to be present, irrespective of the polarization of the incoming helicities, as predicted in Fig.~\ref{fig:SFG_selection} for the $n=1$ case. This is precisely what the results in Table~\ref{tab:rhomboid_800nm_circ} show, as all SHG values have the same order of magnitude.

For the other structure, which shows C\textsubscript{3}-symmetry, the intensity of the SHG wave in transmission and reflection for both plus and minus helicity is shown in Table~\ref{tab:hexagon_800nm}. In the case of equal incoming helicities, the intensity of the second-harmonic signal in transmission (reflection) practically vanishes if the outgoing helicity is the same (opposite) as the incoming helicity. Furthermore, the second row shows that the second-harmonic signal is practically zero in reflection and transmission if the incoming fields have different helicities. These results agree entirely with the selection rules shown in Fig.~\ref{fig:SFG_selection} for $n=3$.

\begin{table}[H]
    \centering
    \begin{tabular}{|c||c|c|c|c|}
        \hline
        \multirow{2}{*}{Incident} & \multicolumn{2}{|c|}{Transm. (arb. units)} & \multicolumn{2}{|c|}{Refl. (arb. units)} \\
        \cline{2-5}
         &     TE & TM&TE &TM \\
        \hline
        TE--TE &9.11e-10   &0.498 & 9.11e-10 &0.498 \\
        \hline
        TM--TM &1.42e-09  &1 & 1.42e-09 & 1 \\
        \hline
        TE--TM &0.383 &2.44e-09 & 0.383& 2.44e-09 \\
        \hline
    \end{tabular}
    \caption{Simulated SHG signal for the rhomboid structure. The fundamental wavelength is 800 nm. The incoming and outgoing polarizations are in the TE/TM basis, and the numbers are normalized to 6.1$\times 10^{-34}$.}
    \label{tab:rhomboid_800nm_lin}
\end{table}

\begin{table}[H]
    \centering
    \begin{tabular}{|c||c|c|c|c|}
        \hline
        \multirow{2}{*}{Incident} & \multicolumn{2}{|c|}{Transm. (arb. units)} & \multicolumn{2}{|c|}{Refl. (arb. units)} \\
        \cline{2-5}
         &$+$ & $-$&$+$ &$-$ \\
        \hline
        $++$ &0.160  &0.254 & 0.254 &0.160 \\
        \hline
        $+-$ &0.359  &0.359 & 0.359 & 0.359 \\
        \hline
    \end{tabular}
    \caption{Simulated SHG signal for the rhomboid structure. The fundamental wavelength is 800 nm. The incoming and outgoing polarizations are in the helicity basis, and the numbers are normalized to 6.1$\times 10^{-34}$.}
    \label{tab:rhomboid_800nm_circ}
\end{table}

\begin{table}[H]
    \centering
    \begin{tabular}{|c||c|c|c|c|}
        \hline
        \multirow{2}{*}{Incident} & \multicolumn{2}{|c|}{Transm. (arb. units)} & \multicolumn{2}{|c|}{Refl. (arb. units)} \\
        \cline{2-5}
         &$+$ & $-$&$+$ &$-$ \\
        \hline
        $++$ & 4.29e-06 &0.139&0.139&4.29e-06 \\
        \hline
        $+-$ &3.67e-05 &3.64e-05 &3.64e-05&3.67e-05 \\
        \hline
    \end{tabular}
    \caption{Simulated SHG signal for the C\textsubscript{3}-symmetric structure. The fundamental wavelength is 800~nm. The incoming and outgoing polarizations are in the helicity basis, and the numbers are normalized to 6.1$\times 10^{-34}$.}
    \label{tab:hexagon_800nm}
\end{table}

\section{The T-matrix in $\M_2$ \label{sec:tmatrix}}
In this section, we investigate when and how it is possible to compute $S_2$ by means of the T-matrix. We start with a short introduction to the T-matrix in $\M$, which is a popular and powerful approach to the computation of light-matter interactions for classical fields and single-photon states \cite{Waterman1965,Gouesbet2019,Mishchenko2020}. In $\M$, the connection between the T-matrix and the scattering operator is:
\begin{equation}
    S=\mathbbm 1 + T.\label{eq:S+T}
\end{equation}
The difference with the typical connection, $S=\mathbbm 1+2T$ \cite{LeRu2013,Mishchenko2020}, is due to the conventions in \cite{vavilin_polychromatic_2024}, which are more appropriate for the polychromatic case.

While it is obvious from \eqref{eq:S+T} that $S$ and $T$ are equivalent regarding their information content, they connect different kinds of fields. While $S$ connects the incoming and outgoing fields, $T$ connects incident fields to outgoing fields. The outgoing field due to $T$ is typically called the scattered field. The incident field is defined as the field that would be present if there was no scattering object, and has both incoming and outgoing components. The total outgoing field is the scattered field plus the outgoing part of the incident field, and the incoming field is the incoming part of the incident field.

While $T$ is the central object in computational techniques, the total outgoing field produced by $S$ is much more relevant in experiments, because the scattered field produced by $T$ is extremely hard to measure independently \cite{Husnik2012}. 

New possibilities appear when moving from $\M$ to $\M_2$. In the interaction of bipartite states of light with matter, we need to distinguish between two different kinds of processes, depending on whether the response of the matter to one part of the state depends on the other part of the state, or not. When it is independent, the process can be modeled by a separable operator in $\M_2$, namely, by the tensor product of two identical operators, each acting separately in one of the two copies of $\M$. When the presence of one part of the state modifies the response of the matter to the other part, the operator for such processes must be non-separable and cannot be written as the tensor product of two operators. Accordingly, we write the total $S_2$ as the sum of a separable operator $S\otimes S$, and a non-separable operator $N_2$
\begin{equation}
     S_2 = S\otimes S+N_2  \label{eq:S2general},
\end{equation}
where $S$ is the scattering operator in $\M$. Mathematically, the non-separability of $N_2$ can be expressed for example in the plane-wave basis as:
\begin{equation}
\label{eq:Npw}
	\begin{split}
		N_{2,\lambda\lambdaBar\sigma\Bar{\sigma}}(\kvec,\kbar,\vecbold{q},\vecbold{\Bar{q}}) &=(\biphotonsimplebra{k}{\lambda})N_2\left(\biphotonsimple{q}{\sigma}\right)\\
		&\neq A_{\lambda\lambdaBar}(\kvec,\kbar)B_{\sigma\Bar{\sigma}}(\vecbold{q},\vecbold{\Bar{q}}).
	\end{split}
\end{equation}
In most situations, $S$ is assumed to preserve the frequency of the fields, which means that an incoming field with a given frequency $\omega_{\text{in}}$ can only produce an outgoing field of the same frequency. The same holds then for the separable part $S\otimes S$. In contrast, the non-separable part $N_2$ allows effects that are absent in $S\otimes S$. For example, one part of the incoming state with frequency $\omega_1$ may induce current densities with the same frequency inside the object, which can then combine with the other part of the incoming state, of frequency $\omega_2$, to produce an outgoing field at the sum frequency $\omega_1+\omega_2$.

Before continuing, it is worth mentioning that the derivation of the selection rules in Sec.~\ref{sec:selectionrules} does not require explicit knowledge of the scattering operator $S_2$. As long as the symmetry condition in Eq.~\eqref{eq:s2sym} is valid, the selection rules follow simply by acting with the symmetry operators on the incoming and outgoing states. Therefore, the selection rules hold for both the separable and non-separable cases. Moreover, while in Sec.~\ref{sec:selectionrules} we have focused on non-linear processes, the selection rules can be obtained with the same procedure for the case of bipartite scattering processes that do not change the frequency. 

\subsection{Separable $S_2$}
It is well-known that incoming fields of large amplitude are needed for measuring the non-linear effects expressed through $N_2$. For example, such effects are negligible in the interaction of entangled biphoton states with nanoparticles, because the incoming fields have particularly weak intensities. In such cases, and many others, one may neglect $N_2$, and write the approximation:
\begin{equation}
     S_2 \approx S\otimes S  \label{eq:S2sep}.
\end{equation}
Let us now study how to computationally approach this case. Expanding \Eq{eq:S2sep} using \Eq{eq:S+T} readily leads to the definition of the T-matrix in $\M_2$ for the separable case:
\begin{equation}
\begin{split}
     S_2 &= \mathbbm1\otimes\mathbbm1 + \mathbbm1\otimes T +  T\otimes\mathbbm1 + T\otimes  T\\
     &\implies T_2=\mathbbm1\otimes T +  T\otimes\mathbbm1 + T\otimes  T.
     \label{eq:S2T2}
     \end{split}
\end{equation}
Therefore, in the separable case, the T-matrix in $\M$ is sufficient for computing the interaction of general bipartite states of light with material objects. For example, \Eq{eq:S2T2} can be used to quantitatively predict how the entanglement of biphoton states changes upon interaction with a nanostructure. Qualitatively, it is readily shown that if the incoming bipartite state is separable, and $S_2$ is also separable, then the outgoing bipartite state must be separable as well. The case of separable $S_2$ and entangled illumination has been considered in \cite{PhysRevLett.121.173901,Lasa-Alonso_2020}. Each of the terms in $T_2$ in \Eq{eq:S2T2} has an intuitive physical meaning: The first two terms describe the interaction of only one of the two parts of the incident bipartite states, and the third term describes the interaction of both parts.

The definition of $T_2$ in \Eq{eq:S2T2} disagrees with existing literature, for example \cite{schotland_scattering_2016}, where only the last term $ T\otimes T$ is used as $ T_2$. However, the following argument is rather definitively in favor of the definition in \Eq{eq:S2T2}. Let us assume that the scattering is lossless in $\M$. This implies that $S$ must be unitary, and hence 
\begin{equation}
     S^\dagger S = \mathbbm 1 \Leftrightarrow  T +  T^\dagger +   T^\dagger T = 0,
\end{equation}
which is the condition in \cite[Eq.~(9)]{LeRu2013}, or \cite[Eq.~(5.59)]{Mishchenko2002}, except that $S=\mathbbm 1+T$ is used here and $S=\mathbbm 1+2T$ is used in these references.

If the scattering is lossless in $\M$, then the bipartite scattering should also be lossless. Only the definition in Eq.~\eqref{eq:S2T2} ensures that, given that $S$ is unitary, $S_2$ is unitary at its turn. That is: $ S_2^\dagger  S_2 =(S^\dagger\otimes S^\dagger)(S\otimes S)= \mathbbm1\otimes\mathbbm1$. 

\subsection{Non-separable $S_2$}
The presence of $N_2$ in the general definition of $S_2$ in \Eq{eq:S2general}, together with the vacuum state, and the emergent non-linearity explained in Sec.~\ref{sec:M2S2}, provide the appropriate framework for computations of non-linear effects. Devising computational strategies for obtaining $N_2$ for general objects is a subject of future research. Here we discuss some relevant points for guiding such research.

While both SFG and SPDC can be fitted in $N_2$, let us, for the sake of discussion, consider a scattering amplitude for the SFG process:
\begin{equation}
	\begin{split}
		&\left(\bra{\kvec_3\,\lambda_3}\bra{0}\right) S_2\left(\ket{\kvec_1\,\lambda_1}\ket{\kvec_2\,\lambda_2}\right)=\\
		&	\left(\bra{\kvec_3\,\lambda_3}\bra{0}\right) N_2\left(\ket{\kvec_1\,\lambda_1}\ket{\kvec_2\,\lambda_2}\right),
	\end{split}
     \label{eq:AS2}
\end{equation}
where $|\kvec_3|=|\kvec_1|+|\kvec_2|$, and the equality follows because $S\otimes S$ cannot change the frequency.

We recall that the physical meaning of the non-separability is that the response of the object to one part of the state depends on the other part. We may then interpret the SFG process as follows. As a first step, one of the two incident fields perturbs the object and excites current sources with frequency $c|\kvec_1|$ or $c|\kvec_2|$, respectively. The interaction of the second field is described by the \emph{T-matrix of the object after perturbation by a first field}, which is different from $T$. This interpretation is consistent with computational strategies for non-linear optical processes in microscopic systems such as molecules \cite[Chap.~3]{Boyd2020}. Such strategies use second-order perturbation theory, where the second-order term is obtained as a perturbation of an already perturbed system, and the first and second perturbations are due to the first and second parts of a bipartite state, respectively. The resulting response functions are called non-linear molecular hyper-polarizabilities. They can be mapped into corresponding Hyper-T-matrices \cite{https://doi.org/10.1002/adma.202311405}. Then, the non-linear response of some macroscopic objects built out of large ensembles of microscopic systems, such as molecular crystals, can be obtained. However, the Hyper-T-matrix cannot be equal to $N_2$ in general. While $N_2$ has four sets of indexes, as can be seen in \Eq{eq:Npw}, the Hyper-T-matrix and the hyper-polarizabilities have only three sets of indexes. Nevertheless, the appearance of the vacuum state in \Eq{eq:AS2} strongly suggests that $N_{2,\lambda 0 \sigma\Bar{\sigma}}(\kvec,0,\vecbold{q},\vecbold{\Bar{q}})$ and the Hyper-T-matrix are equivalent in some way.

More generally, one may ask whether it is possible to obtain $N_2$ for a general macroscopic object. The time-dependent perturbation induced by the first field suggests that this question could be addressed with the tools that are being developed to treat light-matter interaction in time-varying systems \cite{Ptitcyn2023}. Note that, if successful, such an approach would include the SH signal generated on the interfaces between different materials. This latter effect happens in addition to the SH signal produced in the bulk of a given material, which, to a first approximation, is only possible if the unit cell of the material is not centrosymmetric. While the surface SHG of molecular films has been recently simulated {\em ab initio} \cite{Zerulla2024}, the theoretical approach to surface SHG for macroscopic objects, such as a silicon disk, for instance, is to add a phenomenological tensor on the surface \cite{Dadap:04}.

\section{Conclusion and outlook\label{sec:conc}}
In this article, we introduce a theoretical and computational framework for studying the interaction of material structures with bipartite states of light. The framework allows one to readily derive symmetry-induced selection rules for the general case, that is, for states that can be entangled or non-entangled, and material responses that can be separable or non-separable. In non-separable responses, one of the parts of the bipartite state changes the response of the matter to the other part of the state, giving rise to non-linear effects. When the non-linear effects can be neglected, we obtain $T_2$, the T-matrix for bipartite states, as a simple function of the typical T-matrix for single photons. Then, the change in the entanglement of biphoton states interacting with nanostructures can be readily computed. We highlight that the framework allows one to treat general polychromatic fields, such as pulses of entangled photons, for example. 

The results in this article motivate further research, in particular, to achieve the full integration of non-linear processes in the formalism by means of a non-separable operator, as discussed in Sec.~\ref{sec:tmatrix}. Additionally, the invariant properties of the scalar product can be used to formalize projective measurements for bipartite states, by following steps similar to those taken in \cite[Sec.~III]{FerCor2022b} for the single-photon case. For some computations, it will be beneficial to obtain the expressions of the bipartite wavefunction and the scalar product in the multipolar basis, because T-matrices are often obtained in such basis.

We foresee that the algebraic formalism will become a useful tool for the study and engineering of the interaction of bipartite states of light with nanostructures, and arrangements thereof.\\

\begin{acknowledgements}
L.F. wishes to acknowledge the support from the Jane and Aatos Erkko Foundation under the ``DoinQTech'' project. M.K. and C.R. acknowledge support by the Deutsche Forschungsgemeinschaft (DFG, German Research Foundation) under Germany’s Excellence Strategy via the Excellence Cluster 3D Matter Made to Order (EXC-2082/1-390761711) and from the Carl Zeiss Foundation via the CZF-Focus@HEiKA Program. M.K., C.H., and C.R. acknowledge funding by the Volkswagen Foundation. I.F.C. and C.R. acknowledge support by the Helmholtz Association via the Helmholtz program ``Materials Systems Engineering'' (MSE). B.Z. and C.R. acknowledge support by the KIT through the ``Virtual Materials Design'' (VIRTMAT) project. M.K. and C.R. acknowledge support by the state of Baden--Württemberg through bwHPC and the German Research Foundation (DFG) through grant no. INST 40/575-1 FUGG (JUSTUS 2 cluster) and the HoreKa supercomputer funded by the Ministry of Science, Research and the Arts Baden--Württemberg and by the Federal Ministry of Education and Research. 
C.R. acknowledges support by the German Research Foundation under Grant No. RO 3640/14-1 within Project No. 465163297.
\end{acknowledgements}
\bibliography{references2}

\section{Supplementary Information}
\label{sec:supplement}

\subsection{Details of the Density Functional Theory (DFT) calculations}

A single layer of MoS\textsubscript{2} crystal was extracted from the crystallography file given below. The two distinctive finite-size models were cut out in the shape of an approximate hexagon and rhomboid. Single point energy calculations of those models were performed using the development version of TURBOMOLE 7.8\cite{TURBOMOLE2023}, followed by calculation of linear polarizability tensors at fundamental and second-harmonic frequency and first hyperpolarizability tensors. A PBE0\cite{adamoReliableDensityFunctional1999, ernzerhofAssessmentPerdewBurke1999} hybrid exchange-correlation (XC) density functional has been used in combination with def2-TZVP basis set\cite{weigendBalancedBasisSets2005, weigendAccurateCoulombfittingBasis2006} for all atoms. Molybdenum atoms had Stuttgart relativistic effective core potential to reduce the number of explicitly treated electrons by replacing the core electrons by a potential. A numerical grid size was set to "3". The damping was 0.05 eV for the Lorentzian line-shape profile at half-width at half-maximum (HWHM). To speed up the calculations, the 
multipole-accelerated resolution-of-the-identity (MARI-J)\cite{eichkornAuxiliaryBasisSets1995, eichkornAuxiliaryBasisSets1997,sierkaFastEvaluationCoulomb2003} and semi-numerical approach for the exchange (senex, esenex)\cite{Holzer2020} have been employed.
\\
\\
MoS\textsubscript{2} cif file used to extract finite-size molecular models:\\\\
\# generated using pymatgen\\
data\_MoS2\\
\_symmetry\_space\_group\_name\_H-M   'P 1'\\
\_cell\_length\_a   3.19031570\\
\_cell\_length\_b   3.19031570\\
\_cell\_length\_c   14.87900400\\
\_cell\_angle\_alpha   90.00000000\\
\_cell\_angle\_beta   90.00000000\\
\_cell\_angle\_gamma   120.00000000\\
\_symmetry\_Int\_Tables\_number   1\\
\_chemical\_formula\_structural   MoS2\\
\_chemical\_formula\_sum   'Mo2 S4'\\
\_cell\_volume   131.15106251\\
\_cell\_formula\_units\_Z   2\\
loop\_\\
\indent \_symmetry\_equiv\_pos\_site\_id\\
\indent \_symmetry\_equiv\_pos\_as\_xyz\\
\indent  1  'x, y, z'\\
loop\_\\
\indent \_atom\_type\_symbol\\
\indent \_atom\_type\_oxidation\_number\\
\indent  Mo4+  4.0\\
\indent  S2-  -2.0\\
loop\_\\
\indent \_atom\_site\_type\_symbol\\
\indent \_atom\_site\_label\\
\indent \_atom\_site\_symmetry\_multiplicity\\
\indent \_atom\_site\_fract\_x\\
\indent \_atom\_site\_fract\_y\\
\indent \_atom\_site\_fract\_z\\
\indent \_atom\_site\_occupancy\\
\indent  Mo4+  Mo0  1  0.33333333  0.66666667  0.25000000  1\\
\indent  Mo4+  Mo1  1  0.66666667  0.33333333  0.75000000  1\\
\indent  S2-  S2  1  0.66666667  0.33333333  0.35517400  1\\
\indent  S2-  S3  1  0.33333333  0.66666667  0.85517400  1\\
\indent  S2-  S4  1  0.66666667  0.33333333  0.14482600  1\\
\indent  S2-  S5  1  0.33333333  0.66666667  0.64482600  1\\
\\
Cartesian coordinates of both molecular models:\\
Approximate hexagon:\\
55\\
\\
Mo     -0.00000      6.38063     -0.00000\\
Mo     -0.00000      4.78547      2.76291\\
Mo     -0.00000      3.19032      5.52579\\
Mo     -0.00000      4.78547     -2.76289\\
Mo     -0.00000      3.19032     -0.00000\\
Mo     -0.00000      1.59516      2.76291\\
Mo     -0.00000     -0.00000      5.52579\\
Mo     -0.00000      3.19032     -5.52579\\
Mo     -0.00000      1.59516     -2.76289\\
Mo     -0.00000     -0.00000     -0.00000\\
Mo     -0.00000     -1.59516      2.76291\\
Mo     -0.00000     -3.19032      5.52579\\
Mo     -0.00000     -0.00000     -5.52579\\
Mo     -0.00000     -1.59516     -2.76289\\
Mo     -0.00000     -3.19032     -0.00000\\
Mo     -0.00000     -4.78547      2.76291\\
Mo     -0.00000     -3.19032     -5.52579\\
Mo     -0.00000     -4.78547     -2.76289\\
Mo     -0.00000     -6.38063     -0.00000\\
S      -1.56488      6.38063      1.84193\\
S      -1.56488      4.78547      4.60483\\
S      -1.56488      4.78547     -0.92096\\
S      -1.56488      3.19032      1.84193\\
S      -1.56488      1.59516      4.60483\\
S      -1.56488      3.19032     -3.68386\\
S      -1.56488      1.59516     -0.92096\\
S      -1.56488     -0.00000      1.84193\\
S      -1.56488     -1.59516      4.60483\\
S      -1.56488      1.59516     -6.44675\\
S      -1.56488     -0.00000     -3.68386\\
S      -1.56488     -1.59516     -0.92096\\
S      -1.56488     -3.19032      1.84193\\
S      -1.56488     -4.78547      4.60483\\
S      -1.56488     -1.59516     -6.44675\\
S      -1.56488     -3.19032     -3.68386\\
S      -1.56488     -4.78547     -0.92096\\
S      -1.56488     -6.38063      1.84193\\
S       1.56488      6.38063      1.84193\\
S       1.56488      4.78547      4.60483\\
S       1.56488      4.78547     -0.92096\\
S       1.56488      3.19032      1.84193\\
S       1.56488      1.59516      4.60483\\
S       1.56488      3.19032     -3.68386\\
S       1.56488      1.59516     -0.92096\\
S       1.56488     -0.00000      1.84193\\
S       1.56488     -1.59516      4.60483\\
S       1.56488      1.59516     -6.44675\\
S       1.56488     -0.00000     -3.68386\\
S       1.56488     -1.59516     -0.92096\\
S       1.56488     -3.19032      1.84193\\
S       1.56488     -4.78547      4.60483\\
S       1.56488     -1.59516     -6.44675\\
S       1.56488     -3.19032     -3.68386\\
S       1.56488     -4.78547     -0.92096\\
S       1.56488     -6.38063      1.84193\\
\\
Rhomboid:\\
27\\
\\
Mo     -0.00000      3.19031     -0.73794\\
Mo     -0.00000      1.59515      2.02495\\
Mo     -0.00000      0.00000      4.78785\\
Mo     -0.00000      1.59515     -3.50084\\
Mo     -0.00000      0.00000     -0.73794\\
Mo     -0.00000     -1.59516      2.02495\\
Mo     -0.00000      0.00000     -6.26373\\
Mo     -0.00000     -1.59516     -3.50084\\
Mo     -0.00000     -3.19032     -0.73794\\
S      -1.56488      3.19031      1.10399\\
S      -1.56488      1.59515      3.86688\\
S      -1.56488      0.00000      6.62978\\
S      -1.56488      1.59515     -1.65891\\
S      -1.56488      0.00000      1.10399\\
S      -1.56488     -1.59516      3.86688\\
S      -1.56488      0.00000     -4.42180\\
S      -1.56488     -1.59516     -1.65891\\
S      -1.56488     -3.19032      1.10399\\
S       1.56488      3.19031      1.10399\\
S       1.56488      1.59515      3.86688\\
S       1.56488      0.00000      6.62978\\
S       1.56488      1.59515     -1.65891\\
S       1.56488      0.00000      1.10399\\
S       1.56488     -1.59516      3.86688\\
S       1.56488      0.00000     -4.42180\\
S       1.56488     -1.59516     -1.65891\\
S       1.56488     -3.19032      1.10399\\
\\
An example of the "control" file for the calculation of the polarizabilities at 800 and 400 nm:\\
\\
{\$}title\\
{\$}symmetry c1\\
{\$}coord    file=coord\\
{\$}optimize\\
\indent internal   off\\
\indent redundant  off\\
\indent cartesian  on\\
{\$}atoms\\
\indent    basis =def2-TZVP\\
\indent    jbas  =def2-TZVP\\
mo 1-19\\
\indent    ecp   =mo def2-ecp\\
{\$}basis    file=basis\\
{\$}ecp    file=basis\\
{\$}scfmo   file=mos\\
{\$}scfiterlimit     1000\\
{\$}scfdamp   start=15.000  step=0.050  min=1.000\\
{\$}scfdump\\
{\$}scfdiis\\
\indent maxiter=20\\
{\$}maxcor    1000 MiB  per\_core\\
{\$}scforbitalshift  automatic=.1\\
{\$}energy    file=energy\\
{\$}grad    file=gradient\\
{\$}dft\\
\indent    functional   pbe0\\
\indent    gridsize   3\\
{\$}scfconv   7\\
{\$}denconv 1.0d-7\\
{\$}scftol 1.0d-16\\
{\$}rundimensions\\
\indent   natoms=55\\
{\$}closed shells\\
\indent a       1-421                                  ( 2 )\\
{\$}ricore    16000\\
{\$}rij\\
{\$}jbas    file=auxbasis\\
{\$}marij\\
{\$}escfiterlimit 2000\\
{\$}scfinstab dynpol nm\\
\indent800\\
\indent400\\
{\$}mgiao\\
{\$}magnetic\_response\\
{\$}damped\_response 0.05 eV\\
{\$}rpacor 100000\\
{\$}senex\\
\indent  gridsize 1\\
{\$}esenex\\
{\$}last step     define\\
{\$}end\\
\\
An example of the "control" file for the calculation of the first hyperpolarizabilities at 800 nm:\\
\\
{\$}title\\
{\$}symmetry c1\\
{\$}coord    file=coord\\
{\$}optimize\\
\indent internal   off\\
\indent redundant  off\\
\indent cartesian  on\\
{\$}atoms\\
\indent    basis =def2-TZVP\\
\indent    jbas  =def2-TZVP\\
mo 1-19\\
\indent    ecp   =mo def2-ecp\\
{\$}basis    file=basis\\
{\$}ecp    file=basis\\
{\$}scfmo   file=mos\\
{\$}scfiterlimit     1000\\
{\$}scfdamp   start=15.000  step=0.050  min=1.000\\
{\$}scfdump\\
{\$}scfdiis\\
\indent maxiter=20\\
{\$}maxcor    1000 MiB  per\_core\\
{\$}scforbitalshift  automatic=.1\\
{\$}energy    file=energy\\
{\$}grad    file=gradient\\
{\$}dft\\
\indent    functional   pbe0\\
\indent    gridsize   3\\
{\$}scfconv   7\\
{\$}denconv 1.0d-7\\
{\$}scftol 1.0d-16\\
{\$}rundimensions\\
\indent   natoms=55\\
{\$}closed shells\\
\indent a       1-421                                  ( 2 )\\
{\$}ricore    16000\\
{\$}rij\\
{\$}jbas    file=auxbasis\\
{\$}marij\\
{\$}escfiterlimit 2000\\
{\$}scfinstab hyperpol nm\\
\indent800\\
{\$}mgiao\\
{\$}damped\_response 0.05 eV\\
{\$}rpacor 100000\\
{\$}senex\\
\indent  gridsize 1\\
{\$}esenex\\
{\$}last step     define\\
{\$}end\\

\end{document}